\documentclass[preprint,nofootinbib,amsmath,prd,aps,superscriptaddress,tightenlines,12pt]{revtex4}
 \pdfoutput=1
\usepackage{graphicx}
\usepackage{graphics}
\usepackage{amsmath}
\usepackage{hyperref}

\def\lsim{\mathrel{\rlap{\lower4pt\hbox{\hskip1pt$\sim$}}
    \raise1pt\hbox{$<$}}}                
\def\gsim{\mathrel{\rlap{\lower4pt\hbox{\hskip1pt$\sim$}}
    \raise1pt\hbox{$>$}}}                

\def\OMIT#1{}

\newcommand{\be}{\begin{eqnarray}}
\newcommand{\ee}{\end{eqnarray}}

\newcommand{\nn}{\nonumber}
\newcommand{\w}{\omega}

\newcommand{\bea}{\begin{eqnarray}}
\newcommand{\eea}{\end{eqnarray}}

\def\lsim{\mathrel{\rlap{\lower4pt\hbox{\hskip1pt$\sim$}}
    \raise1pt\hbox{$<$}}}                
\def\gsim{\mathrel{\rlap{\lower4pt\hbox{\hskip1pt$\sim$}}
    \raise1pt\hbox{$>$}}}                

\def\OMIT#1{}

\begin{document}


\title{\bf  High precision predictions for exclusive $VH$ production at the LHC}

\author{Ye Li}
\email{yli@slac.stanford.edu}

\affiliation{SLAC National Accelerator Laboratory,
                   Stanford University, Stanford,
                    CA 94309}
                   
\author{Xiaohui Liu}
\email{xiaohuiliu@anl.gov}

\affiliation{High Energy Division, 
                  Argonne National Laboratory, 
                  Argonne, IL 60439}
\affiliation{Department of Physics and Astronomy, 
                   Northwestern University,
                   Evanston, IL 60208}



\newpage
\preprint{SLAC-PUB-15884}


\begin{abstract}
  \vspace*{0.3cm}

We present a resummation-improved prediction for $pp\to VH + 0$ jets at the Large 
Hadron Collider. We focus on highly-boosted final states in the presence of jet veto to suppress
the $t{\bar t}$ background.
In this case, conventional 
fixed-order calculations are plagued by the existence of large Sudakov logarithms 
$\alpha_s^n \log^m (p_T^{veto}/Q)$ for $Q\sim m_V + m_H$ which lead to unreliable predictions as well
as large theoretical uncertainties, and thus limit the accuracy when comparing experimental measurements to the Standard Model.
In this work, we show that the resummation of Sudakov logarithms beyond the
next-to-next-to-leading-log accuracy, combined with the  
next-to-next-to-leading order calculation, reduces the scale uncertainty and stabilizes the perturbative expansion in the region where the vector bosons carry large transverse momentum.
Our result improves the precision with which Higgs properties can be determined from LHC measurements using boosted Higgs techniques.
\end{abstract}

\maketitle

\setlength\baselineskip{17pt}

\section{Introduction}
With the confirmation of the existence of a Higgs particle by both the CMS and ATLAS collaborations~\cite{AtlasHiggs,CMSHiggs} at the Large Hadron Collider (LHC), one of the main objectives of the LHC is to test as many of its properties as possible. The experimental discovery channel is based on its bosonic decay only, and it becomes crucial to probe its couplings to fermions directly to verify whether they agree with Standard Model predictions. However, the standard Higgs production channel of gluon fusion suffers from large QCD backgrounds in its fermion decay modes. The so-called Higgsstrahlung process provides another important way of Higgs measurement at hadron colliders, where the Higgs is produced in association with either a $W$ or $Z$ vector boson (we will refer to them generally as $V$). It was the main search channel for a light Higgs at the Tevatron. At the LHC, the large background from the semi-leptonic decay of the $t\bar{t}$ process obscures the Higgs signal and leads to a poor signal to background ratio. However, recent studies~\cite{Butterworth:2008iy} have shown that with techniques such as jet substructure and large transverse momentum cuts on the Higgs and the vector boson, the Higgsstrahlung process offers a viable  alternative to study the fermion decay channel~\cite{Chatrchyan:2013zna,Chatrchyan:2012ww,TheATLAScollaboration:2013lia,Aad:2012gxa} as well as possible invisible decays of the Higgs~\cite{CMS:2013yda,ATLAS:2013pma}. To further suppress the unwanted backgrounds, both CMS and ATLAS adopt a jet veto procedure allowing no extra jets with $p_T$ greater than $p_T^{veto} \sim 25{\rm GeV}$.

Though effective in the experimental analysis, on the theory side, imposing a jet veto with $p_T^{veto}$ much less than the center of mass energy at which the hard processes take place
leads to large Sudakov logarithms that can destabilize the perturbative series in conventional perturbative QCD calculations. 
 It has been shown that the impact of QCD corrections is sizable for events with $W$ and $H$ boosted at large $p_T$ in the presence of jet veto\footnote{The imposed jet veto makes the lower $p_T$ cut on the vector boson equivalent to that on the Higgs in the boosted region.} in Ref.~\cite{Ferrera:2011bk}, where  
the fully exclusive cross section 
for $WH$ production up to next-to-next-to-leading order (NNLO) in QCD was studied. The jet veto reverses the sign of the higher-order QCD correction compared to the inclusive case, and reduces the total cross section. Recently there have been works  on the transverse-momentum resummation of $V+H$~\cite{Dawson:2012gs} and on the jet veto resummation~\cite{Shao:2013uba} for the Higgsstrahlung process. Neither study addressed the impact of the jet veto as a function of the transverse momentum of $V$ or $H$, which is required to understand the QCD corrections in the boosted phase-space region.

In this manuscript, we present the fully exclusive cross section for $VH$ production at NNLO in QCD matched with the resummation of jet veto logarithms at the partial next-to-next-to-leading-log prime (${\rm NNLL}'_p$) accuracy and study the relevant theoretical uncertainty. 
 We follow the counting of logarithmic accuracy according to Ref.~{\cite{Stewart:2013faa}}, and the subscript $p$ stands for ``partial", indicating that we have not included the non-logarithmically enhanced contributions at ${\cal O}(\alpha_s^2)$, the effect of which is found to be small. In the 
 small-jet-radius limit, our calculation is expected to be comparable with the ${\rm NNLL}'$ accuracy.
Compared to Ref.~\cite{Shao:2013uba}, we have achieved higher accuracy in both the resummation and the fixed order calculation. In addition, we follow the experimental analysis 
more closely by studying the transverse momentum dependence of the cross section and focusing on the impact of the resummation in the boosted region of $V$ and $H$. 

Early effort in resuming jet veto was first investigated in Ref.~\cite{Berger:2010xi} using the global variable beam thrust. The resummation of jet veto logarithms to ${\rm NNLL}'_p$ here relies on recent developments in the soft collinear effective theory (SCET)~\cite{Bauer:2000ew, Bauer:2000yr, Bauer:2001ct, Bauer:2001yt, Bauer:2002nz}, which has been applied successfully to the $H+0j$ production~\cite{Becher:2012qa,Becher:2013xia,Stewart:2013faa}. Using the effective theory, the $H+1j$ case was studied in Ref.~\cite{Liu:2012sz} and Ref.~\cite{Liu:2013hba} for $p_T^J \sim m_H$ and was extended to the challenging low jet $p_T$ region in Ref.~\cite{Boughezal:2013oha}. A complete framework for the combination of resummed results for production processes in different exclusive jet bins was also presented in Ref.~\cite{Boughezal:2013oha}. A systematic study on the jet veto clustering effects up to ${\cal O}(\alpha_s^3)$ in the small-jet-radius limit was carried out in Ref.~\cite{Alioli:2013hba}.

The NNLO QCD correction is obtained by modifying the numerical code FEWZ~\cite{Gavin:2010az, Gavin:2012sy, Li:2012wna}, originally used to calculate the DY process. The heavy-quark induced process is not included. 
 Higgs production in association with a vector boson is dominated by DY like processes up to next-to-leading order (NLO) in QCD. At NNLO, there are small contribution from the process where the Higgs is produced via heavy quark loop induced by gluon pairs. Its contribution is found to be around 1\%(1\%) for $WH$ and 5\%(9\%) for $ZH$ inclusive cross section for an 8 TeV (14 TeV) LHC~\cite{Brein:2012ne}. This is expected to be further suppressed in the region where the vector boson and Higgs carry large transverse momentum\footnote{for the $ZH$ process, this subleading process actually is enhanced below the top pair threshold, but here we will only focus on the resummation on its DY like contribution for the purpose of this manuscript.}. 
 Furthermore, the decay of the vector boson into lepton final states is available as a result of the nature of the original FEWZ. The modification is validated via a series of numerical checks against MCFM~\cite{Campbell:2010ff} for differential observables at NLO, Sherpa~\cite{Gleisberg:2003xi} for $VH$ plus 1 or 2 jet(s) exclusive results, as well as VH@NNLO~\cite{Brein:2012ne} for inclusive cross sections. In this study, we integrate over the lepton phase space inclusively for better numerical stability. 
We found that the scale uncertainty gets reduced and the convergence of the perturbative series
is improved dramatically after resumming the jet veto logarithms.
The results of our analysis should stay largely unaffected in the presence of standard experimental acceptance cuts due to the absence of new phase space singularities at the given perturbative order. 
   
Our manuscript is organized as follows. In section~\ref{theory}, we briefly review the theoretical set-up in the resummation of jet veto logarithms. In section~\ref{Numerics}, we
present the numerical consequence of the resummation for both $ZH$ and $WH$ production
at the LHC. We adopt cuts similar to the current experimental analyses and concentrate on the highly-boosted region. Finally, we conclude in Section~\ref{concl}. All necessary technical details are given in the Appendix.

\section{Review of the theoretical formalism}\label{theory}
In this section, we briefly review the theoretical framework for understanding jet vetoed cross
sections at hadron colliders. The formalism has been established in a series of works
for both $0$ jet and exclusive $1$ jet based on either the QCD coherence argument~\cite{Banfi:2012yh,Banfi:2012jm} or the effective theory analysis~\cite{Becher:2012qa,Becher:2013xia,Stewart:2013faa,Liu:2012sz,Liu:2013hba,Boughezal:2013oha}. 
Our approach relies on the recently developed works in SCET~\cite{Bauer:2000ew, Bauer:2000yr, Bauer:2001ct, Bauer:2001yt, Bauer:2002nz}.
Here we will only highlight the formulae used in our calculation and
we refer the readers to Refs.~\cite{Becher:2012qa,Becher:2013xia,Stewart:2013faa,Liu:2012sz,Liu:2013hba} for a detailed discussion and derivation of the factorization theorem
 for jet vetoed cross sections and to Refs.~\cite{Campbell:2013qaa, Heinemeyer:2013tqa} for short reviews. We group all the necessary ingredients for 
 ${\rm NNLL}'_p$ resummation in the Appendix.

\subsection{factorization, resummation and matching}\label{factorization}
When $p_T^{veto}$ is much less than the energy scale $Q$ which characterizes the production process, the cross section can be approximated by a factorized piece with an additional term,
\bea\label{xsec}
\frac{\mathrm{d}^2\sigma}{\mathrm{d}Q^2 \mathrm{d}Y} = 
\frac{\mathrm{d}\hat{\sigma}_B}{\mathrm{d}Q^2} \,
H(Q^2,\mu)\,
B\left(x_a,p_T^{veto},R,\mu,\nu \right) \,
B\left(x_b,p_T^{veto},R,\mu,\nu \right)\,
S\left(p_T^{veto}, R, \mu, \nu \right) + \sigma_0^{\rm Rsub} \,,  \nn \\
\eea
 up to contributions suppressed by $p_T^{veto}/Q$, where $R$ is 
 the jet size parameter for the clustering algorithm. Here $\mathrm{d}\hat{\sigma}_B$ is the Born level partonic cross section initiating the 
relevant process. For instance for $ZH$ production the Born level cross section comes from the tree level process
$q{\bar q} \to Z^* \to ZH \to l^+l^- H$ and $q {\bar q} \to WH \to \nu l H$. Here
$Q^2$ and $Y$ are the invariant mass square and rapidity of the entire final state, respectively.
The Bjorken scale parameter $x$ is given by
\bea
x_{a,b} = \frac{Q}{E_{cm}} e^{\pm Y} \, \in (0,1)\,,
\eea
with $E_{cm}$ denoting the machine center of mass energy. 
Here $H$, $B$ and $S$ are the hard function, beam function for describing the collinear radiations in the forward region and soft function for the radiations with low energies. Their
field theoretic definition can be found in Ref~\cite{Becher:2012qa}. Their explicit form up to ${\cal O}(\alpha_s^2)$
is given in the Appendix.

Due to the homogenous expansion in ${\rm SCET}_{\rm II}$ where the soft and the collinear modes share the same scaling in their virtualities, other than the normal 
divergence regulated by the renormalization scale $\mu$ in the dimensional regularization, there exists additional rapidity divergences which are regulated
by a new fictitious scale $\nu$ in both the beam and soft functions~\cite{Chiu:2011qc, Chiu:2012ir}. The regularization scheme gives rise
to a new renormalization group equation in SCET, which eventually allows us to sum up
the full set of large logarithms of the form $\log(p_T^{veto}/Q)$ to all orders, by evolving 
each function in Eq.~(\ref{xsec}) from its natural scales 
$(\mu_i,\nu_i)$
to the common scales $(\mu,\nu)$. The scales $(\mu_i,\nu_i)$ are estimated
by demanding that in each function the perturbative series behaves properly. Therefore, we have
\bea
&&\mu_H \sim -iQ, \quad \mu_B \sim \mu_S \sim p_T^{veto} \,, \nn \\
&&\nu_B \sim Q, \quad \quad \nu_s \sim p_T^{veto} \,.
\eea
Here choosing the imaginary scale $\mu_H \sim -iQ$ allows us to sum up a tower of large $\pi^2$ terms
for time-like processes~\cite{Ahrens:2008qu}. 

Eq.~(\ref{xsec}) also contains a non-factorizable correction term, 
\bea
\sigma_0^{\rm Rsub}  \propto \left( \frac{\alpha_sC_F}{4\pi}\right)^2\,
\left( 
\frac{16\pi^2}{3}R^2 - 4 R^4
\right) \log\frac{p_T^{veto}}{Q} \,,
\eea
which contributes at the ${\rm NNLL'}_p$ level especially when $R\sim 1$. As argued in 
Refs~\cite{Banfi:2012yh,Banfi:2012jm},  the overall coefficient 
determines the complete logarithmic series coming from this
term. Therefore, we follow the procedure in Ref.~\cite{Stewart:2013faa} to include this piece in
the resummed cross section at ${\rm NNLL'}_p$ by multiplying it with the total evolution factor.

The SCET cross section Eq.~(\ref{xsec}) only gives the exact result when $p_T^{veto}/Q \to 0$. When away from zero, the missing power-suppressed term may have a sizable 
effect on the cross section. In order to recover its contribution, we have to match 
the SCET calculation onto the fixed-order QCD cross section. Here we adopt the most straightforward matching scheme, in which we subtract the most singular terms in $p_T^{veto}/Q$ predicted by SCET up to two loops, from the NNLO QCD cross section, and replace
that by the ${\rm NNLL}'_p$ along with $\pi^2$ resummed one (see, section~\ref{Numerics}). The final result for the resummation-improved cross section is 
\bea\label{matching}
\sigma_{{\rm NNLL'}_p+{\rm NNLO}} = 
\sigma _{\rm NNLO} - \sigma^{\rm sing.}_{{\rm NNLL'}_p} + \sigma^{\rm resum}_{{\rm NNLL'}_p}  \,.
\eea
Here $\sigma _{\rm NNLO}$ is the NNLO cross section obtained based on a 
modification of the FEWZ code, detailed in Appendix~\ref{sec:fewzh}. The SCET expanded cross section  
$\sigma^{\rm sing.}_{{\rm NNLL'}_p} $
 contains all the ${\cal O}(\alpha_s^2)$ order 
singular contributions to $\sigma _{\rm NNLO}$,
and $\sigma^{\rm resum}_{{\rm NNLL}'_p} $ denotes the cross section for with ${\rm NNLL}'_p + \pi^2$  resummation. All the
building blocks needed for $\sigma^{\rm sing.}_{{\rm NNLL'}_p} $ and 
 $\sigma^{\rm resum}_{{\rm NNLL'}_p} $ can be found in the Appendix.

\subsection{$\log(R)$ dependent contributions}
For a complete $\text{NNLL}'$ $p_{T}^{veto}$-resummation, all the matrix elements 
should be computed up to ${\cal O}(\alpha_s^2)$ (see for instance, Ref.~\cite{Stewart:2013faa}). Starting from the $\alpha_s^2$ order, clustering
effects will give rise to $\log(R)$ dependent contributions, which, for small jet radius, dominate over the other non-$\log(p_T^{veto}/\mu)$  enhanced terms of 
the two loop
matrix elements. The 
$\log(R)$ corrections at $\alpha_s^2$
appearing in both the soft and beam functions comes from the clustering of two correlated radiations coming from a single collinear splitting
of distance roughly $R$ into two different jets. They can therefore be extracted from
the strongly ordered collinear behavior of QCD, given by
\bea
\Delta |{\cal M}|^2_{jk,\rm NNLO} = \int [\mathrm{d} k]\,
\mathrm{d}x \,
|{\cal M}|_{i,\rm NLO}^2(k_T,z,\mu,\nu,\epsilon,\eta)
\int \,
\frac{\mathrm{d}q_\perp^2}{q_\perp^2}\,
\frac{\alpha_s}{2\pi}P_{jk\leftarrow i}(x) \,
\Theta(\Delta R-R)\Delta \hat{{\cal F}}_{veto} \,,\nn \\
\eea
 where, $|{\cal M}_{i,\rm NLO}|^2$ is the one loop soft or beam function with
 properly regularizing the divergence in the effective theory and $P_{jk\leftarrow i}(x)$ is the one
 loop splitting function. The transverse momentum $q_\perp$ is with respect to the mother particle $i$ before splitting and $\Delta \hat{{\cal F}}_{veto} $ is
 the phase space measure which accounts for the jet algorithm. We note 
 that an additional symmetry factor should be included when evaluating 
 $g\to gg $ or $q \to qg $ with the gluon or the quark further splitting, respectively, in the beam function.
 
 Using the fact that in the small $R$ limit,
 \bea
\Delta R^2 \approx \frac{q_\perp^2}{k_T^2z^2(1-z)^2}\,,
\eea
and writing out explicitly the measure  $\Delta \hat{{\cal F}}_{veto} $, we get
\bea
\Delta |{\cal M}|^2_{jk,\rm NNLO} &=& \frac{\alpha_s}{2\pi}
\int [\mathrm{d}k]\,  \mathrm{d}x \,
|{\cal M}|_{i,\rm NLO}^2(k_T,z,\mu,\nu,\epsilon,\eta)\nn \\
&&\hspace{-3.ex}  \times \int _R^{\sim 1} \frac{\mathrm{d}\Delta R^2 }{\Delta R^2}\,
P_{jk\leftarrow i}(x) 
\left[\Theta\left(p_T^{veto}-\max(x,1-x) k_T  \right)
- \Theta \left( p_T^{veto } - k_T\right)
 \right] \,.
\eea
Using the equation above, all the $\alpha_s^2$ order $\log(R) $ terms, 
either in the anomalous dimensions or in the matrix elements,
can be calculated
in a straightforward way. For our purposes, we have the following $p_T^{veto}$ 
independent $\log(R)$ terms in the beam function:
\bea
&&I_{q_iq_i,\log R}^{(2)}  = 
\frac{2}{9}C_F\left( 
\left(-12 \pi ^2+131-132 \log (2)\right) C_A
+  (24 \log (2)-23) n_f T_F
\right)
\log(R^2) \, p_{q_iq_i}^{(0)}(z)\,,
\nn \\
&&I_{q_ig,\log R}^{(2)}  =\,
4 C_FT_F \left(-\frac{\pi ^2}{3}+3-3 \log (2)\right)
  \log(R^2)
 p_{q_i,g}^{(0)}(z)\,,
\eea
where $p^{(0)}_{ij}(z)$ is the splitting function which can be found in the appendix.

We note that other than the $\log(R)$ contributions, we also need the non-$\log(R)$ enhanced
$p_T^{veto}$ independent terms at ${\cal O}(\alpha_s^2)$ for claiming $\text{NNLL}'$ accuracy. 
The analytic formula for the full two loop soft function including these contributions is known~\cite{Stewart:2013faa}, 
and is listed in the appendix. For the beam function, 
those terms can be obtained easily by comparing the fixed two loop SCET expansion with
the full ${\rm NNLO}$ QCD calculation. We found that for $pp\to VH$, these contributions are numerically negligible for both large and small $R$, therefore in this work we do not
include their contributions and denote our accuracy as ${\rm NNLL}'_p$.

\section{Numerical results}\label{Numerics}
We start the section by demonstrating the validity of the SCET 
factorization theorem numerically. Later on, we will introduce our scale choices and our strategy for the theoretical uncertainty
estimation.

\begin{figure}[h!]
\begin{minipage}[b]{3.0in}
  \includegraphics[width=3.0in,angle=0]{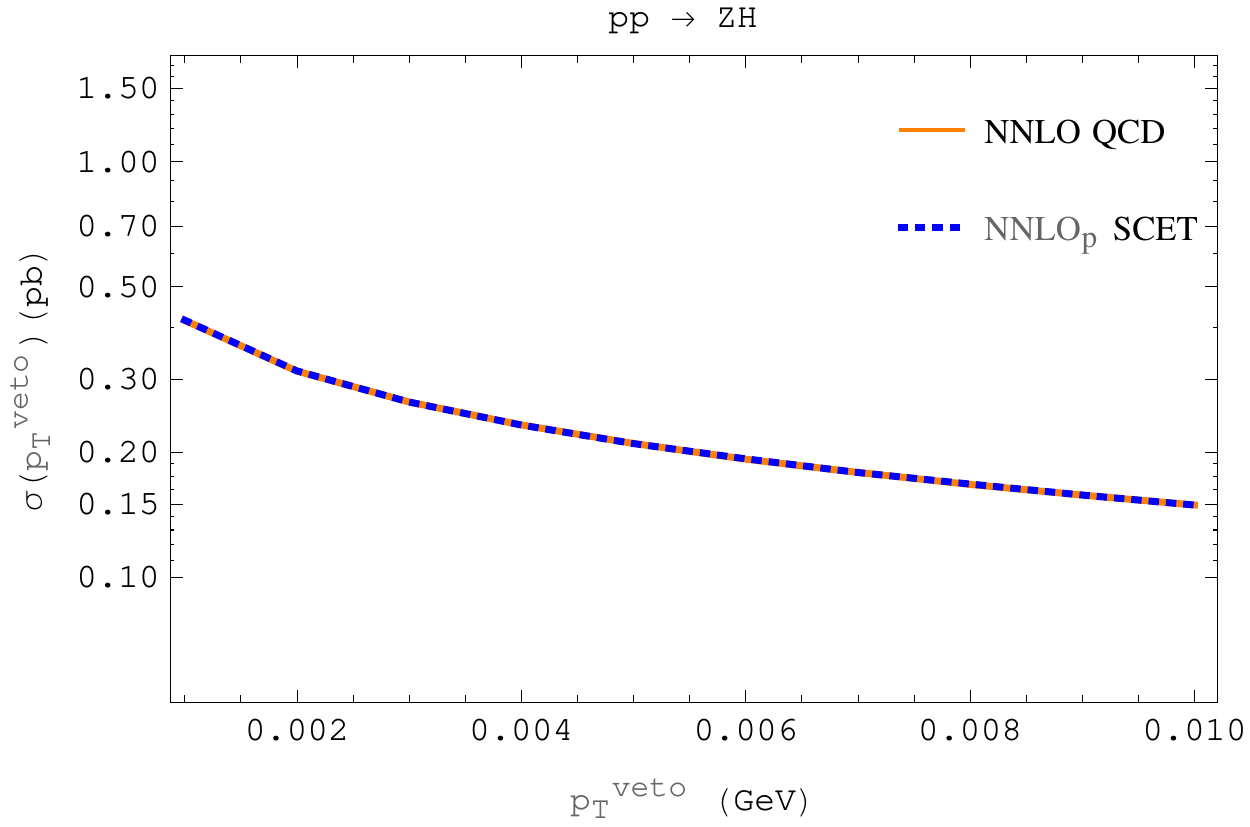}
\end{minipage}
\begin{minipage}[b]{3.0in}
  \includegraphics[width=3.0in,angle=0]{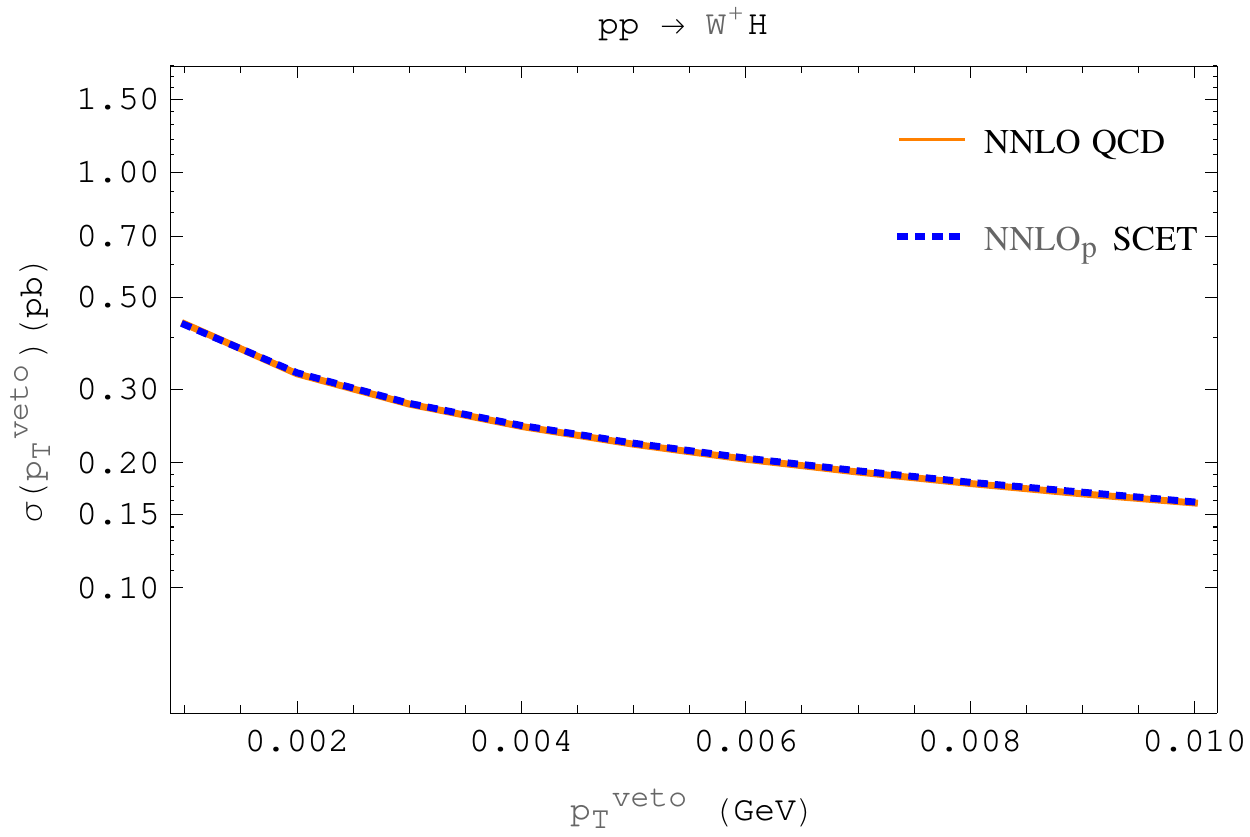}
\end{minipage}
\caption{We compare the NNLO QCD (red solid) cross section 
with the expanded NNLL to ${\cal O} \left(\alpha_s^2\right)$ from
SCET prediction (blue dotted) for both $ZH$ with $p^{T}_{Z} > 100\>{\rm GeV}$ (left panel) 
and $W^+H$ with $p^{T}_{W}>180\>{\rm GeV}$ production (right panel) at LHC14.  
We focus on the small values of $p_T^{veto}$. 
}
\label{fig:comparison}
\end{figure}

In Fig.~\ref{fig:comparison}, we show the comparison of  the cross section
$\sigma\left(p_T^{veto}\right)$ with a jet veto between the QCD prediction and the SCET calculation up to NNLO in $\alpha_s$. 
The results are given as a function of $p_T^{veto}$ for both $ZH$ and $WH$ production at the LHC with 
$14\>{\rm TeV}$ machine energy (LHC14). The jets are constructed using 
anti-k${}_T$ jet algorithm with $R=1.2$.
For small $p_T^{veto}$, the two calculations differ by a non-logarithmically enhanced constant piece (independent of $p_T^{veto}$) at $\mathcal{O}(\alpha_s^2)$, in addition to $p_T^{veto}/Q$ power suppressed
terms. The numerical results presented in Fig.~\ref{fig:comparison} shows almost no differences between
them for small $p_T^{veto}$. The differences are found to be around or less
than per mil level.
%
%

As $p_T^{veto}$ increases, the size of the power-suppressed term 
starts to grow and becomes important when $p_T^{veto}$ is comparable with $Q$. The 
power suppressed term is properly included in our prediction via the matching procedure, Eq.~(\ref{matching}), 
described in the previous section.

The growth of the power suppressed contribution leads to the idea of
 profile scales~\cite{Liu:2013hba,Stewart:2013faa,Ligeti:2008ac,Abbate:2010xh}
 that smoothly turn off resummation and merge with the fixed-order predictions in the high $p_T^{veto}$ region where the effective theory breaks
down and the power suppressed terms dominate. However, in our present analysis, we focus
a fixed $p_T^{veto}$ of $25{\rm GeV}$ and study boosted $VH$ production.
We expect that for such a small $p_T^{veto}$, the singular 
contributions dominate and must be resummed. Therefore instead of using profile scales,
we parametrize the scales using their canonical values in SCET:
\bea
&&\mu = m_V + m_H \,, \nn \\
&&\mu_H = -iQ = -i\sqrt{x_ax_b}E_{\rm com},  \quad \mu_B = \mu_S = p_T^{veto} \,, \nn \\
&&\nu_B = Q, \quad \quad \nu_S = p_T^{veto} \,.
\eea
as our central scale choice. The imaginary hard scale choice $\mu_H$ allows
a better convergent perturbative series in evaluating the hard function and a resummation 
of a towers of $\pi^2$ terms.

We default to 
\bea
\mu_R = \mu_F = \mu = m_V+m_H \,,
\eea
as the central scale for the fixed order calculation.
We vary the scales up and down by a factor of two to estimate the scale uncertainty.
For the fixed order results, we use the ``Stewart-Tackmann-prescription" for the error 
estimation~\cite{Stewart:2011cf}:
\bea\label{fixerror}
\Delta_{0j}^2 = \Delta^2_{\rm tot.} + \Delta^2_{>1j}\,,
\eea
where $\Delta_{\rm tot.}$ and $\Delta_{>1j}$ are the uncertainties for 
total and $1j$ inclusive cross sections, respectively, 
obtained by varying the scale up and down by a factor two. For the resummation improved
predictions, we use
\bea\label{reserror}
\Delta_{0j}^2 = \Delta^2_{\rm coll.} + \Delta^2_{\rm resum}\,,
\eea
where $\Delta_{\rm coll.}$ is computed by varying all the scales in the effective theory collectively. For $\Delta_{\rm resum}$, we vary each scale $\mu_i$ and $\nu_i$ independently and
take the envelope of all the variations.
 
We now turn to discuss the main results of this work. In presenting our numerical results, we use the (N)NLO MSTW2008 PDF set~\cite{Martin:2009iq} for (N)NLO and ${\rm (N)NLL}+{\rm (N)NLO}$ results  and set the value and running of
the strong coupling constant $\alpha_s$  accordingly. We use $n_f = 5$ for the active number of quark flavors, and choose the $G_{\mu}$ scheme as the electroweak input parameters throughout our analysis. We will only discuss $ZH$ and $W^+H$ productions at LHC14 in this manuscript. Jets are formed with anti-$k_T$ jet algorithm\footnote{Here anti-$k_T$ is selected for illustrative purposes only. There is no theoretical
difficulty for us to switch to other $k_T$ type jet algorithms, for instance Cambridge/Aachen, 
and the conclusions of this work will not be affected.} with $R=1.2$~\cite{Butterworth:2008iy} 
and are vetoed if they have transverse momentum greater than $p_T^{veto} = 25~{\rm GeV}$.  
 
\begin{figure}[h!]
\begin{minipage}[b]{3.0in}
  \includegraphics[width=3.0in,angle=0]{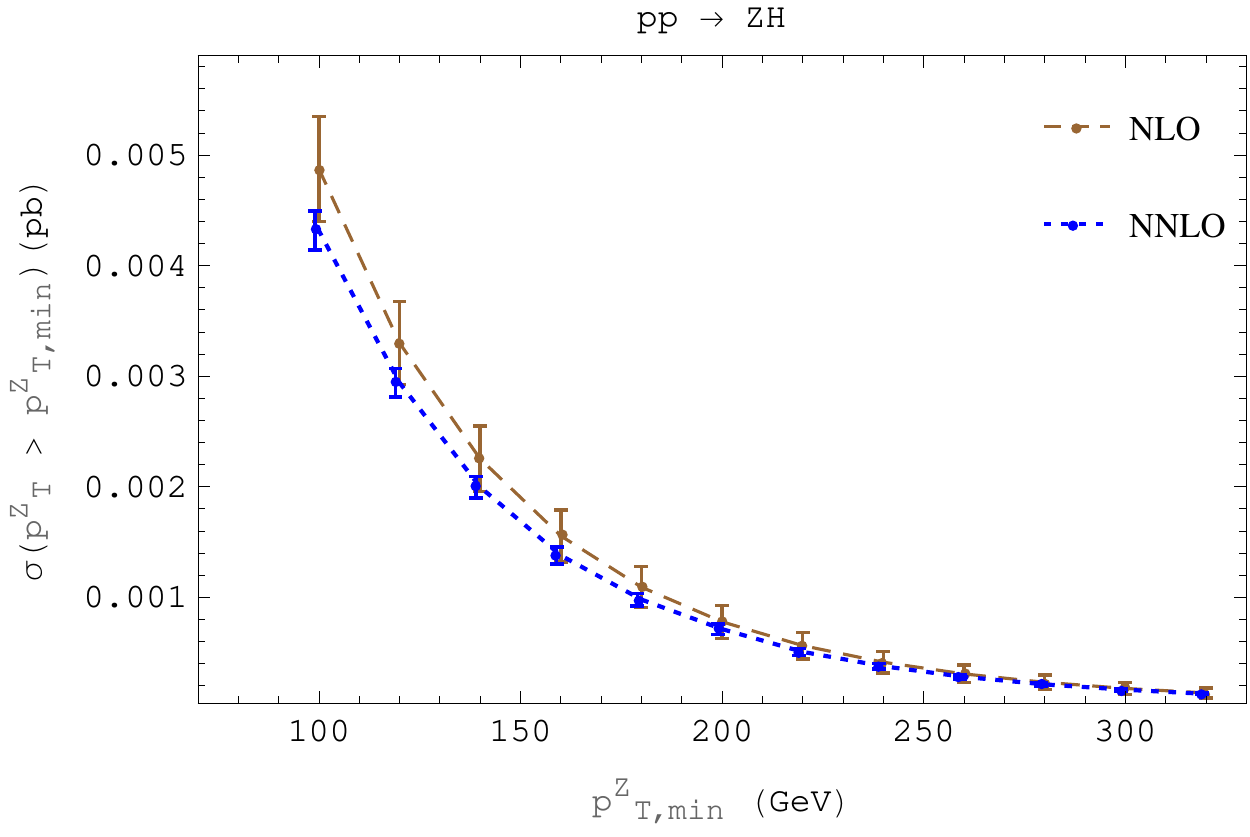}
\end{minipage}
\begin{minipage}[b]{3.0in}
  \includegraphics[width=3.0in,angle=0]{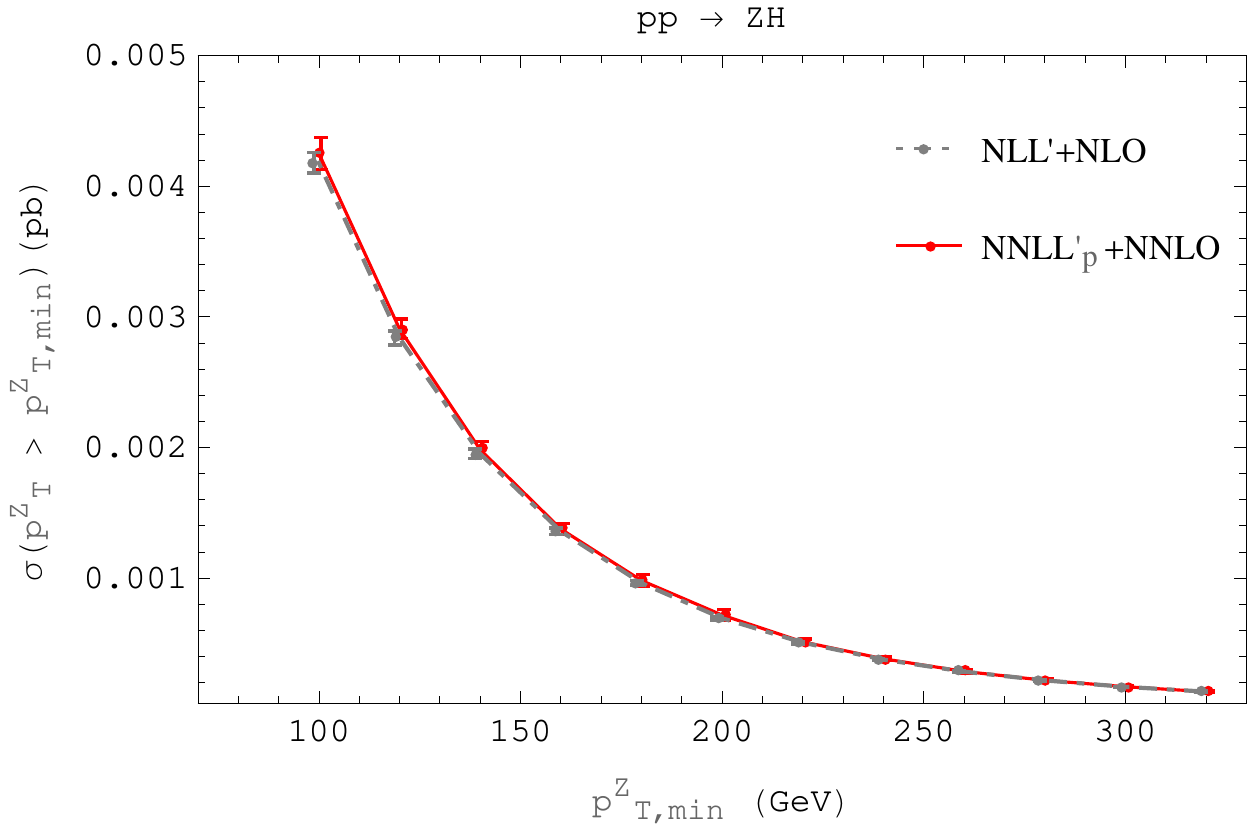}
\end{minipage}
\caption{The NLO (brown dashed) and NNLO (blue dotted) predictions for $ZH$
production at LHC$14$ (left panel) along with the  resummation improved
${\rm NLL}'+{\rm NLO}$ (gray dot-dashed) and ${\rm NNLL}'_p+{\rm NNLO}$ (red solid) results (right panel) are showed, as a comparison of the convergence of the perturbative series with and without the jet veto log resummation. }
\label{fig:zcompare}
\end{figure}

We first study the stability of the perturbative predictions with and without the
improvement from the jet veto resummation.  Fig.~\ref{fig:zcompare} illustrates the impact
of the resummation on the convergence of the  $ZH$ perturbative cross section predictions at LHC$14$. Here we plot 
the cross section as a function of the lower transverse momentum cut of the $Z$ boson $p^Z_{T,{\rm min}}$.
It is clear that the resummation of the jet veto logarithms (right panel) accelerates the convergence of the perturbative series compared to the pure fixed-order ones (left panel),
as the $p_{T,{\rm min}}^Z$ curve from the ${\rm NLL}'+{\rm NLO}$ prediction almost overlaps with the ${\rm NNLL}'_p+{\rm NNLO}$ one, while the NLO prediction is off by a visible amount with respect
to the NNLO result.

\begin{figure}[h!]
\begin{minipage}[b]{3.0in}
  \includegraphics[width=3.0in,angle=0]{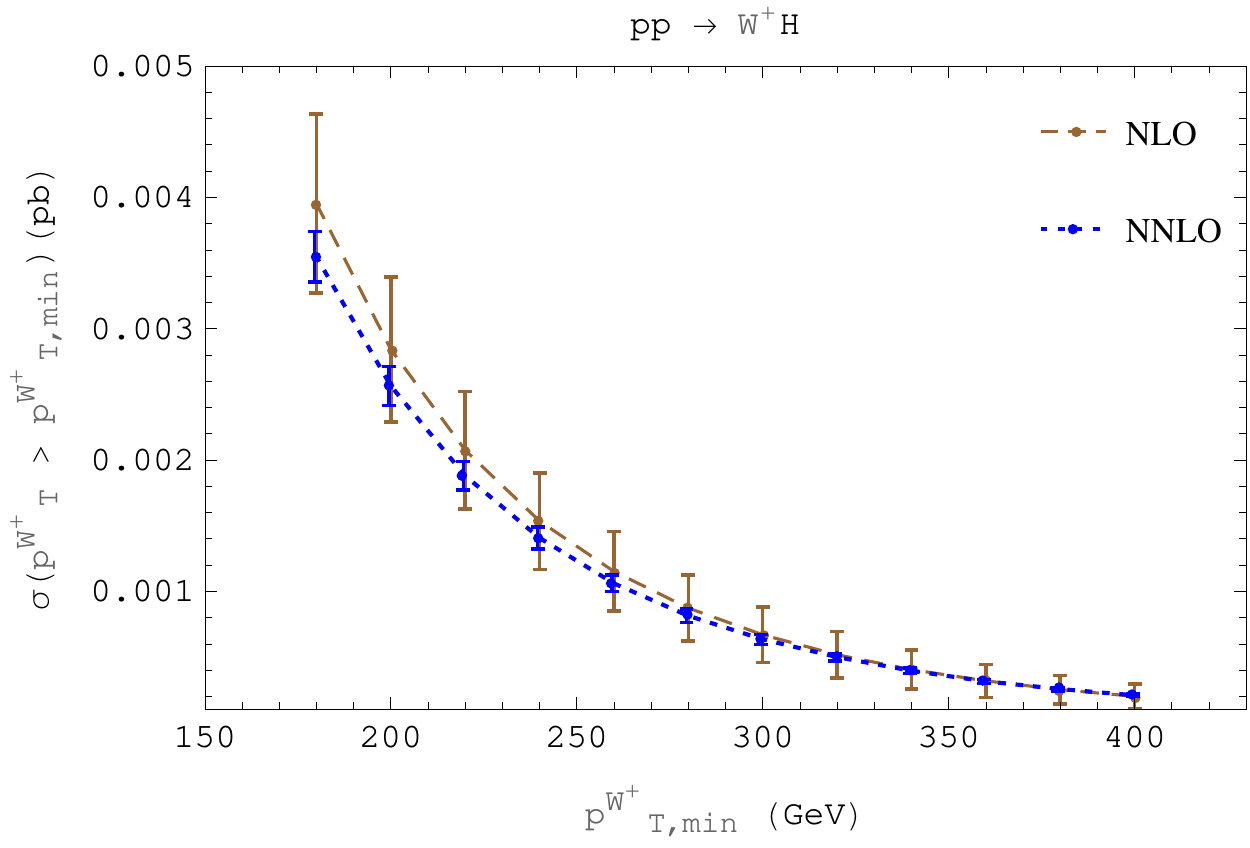}
\end{minipage}
\begin{minipage}[b]{3.0in}
  \includegraphics[width=3.0in,angle=0]{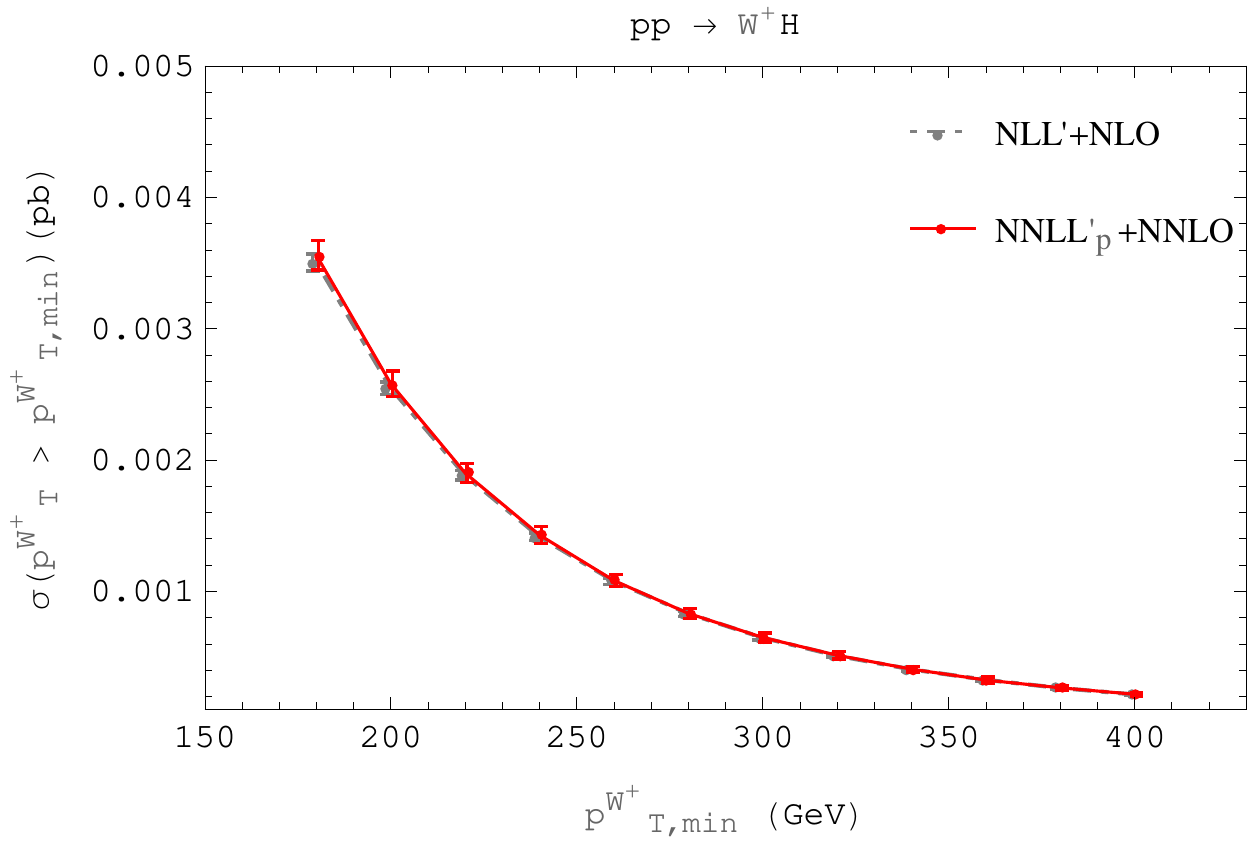}
\end{minipage}
\caption{Similar plots as fig.~\ref{fig:zcompare} for the convergence study on $W^+H$ production
at LHC$14$. The error bars denote the estimated theoretical scale uncertainties. }
\label{fig:wcompare}
\end{figure}

The same trends can be observed
for the predictions of the $W^+H$ production at LHC$14$ which are depicted in fig.~\ref{fig:wcompare}.
The convergence of the perturbative expansion is greatly improved after the resummation.

In both fig.~\ref{fig:zcompare} and fig.~\ref{fig:wcompare}, the error bands reflect
the scale uncertainties estimated using the prescriptions sketched previously in
Eq.~(\ref{fixerror}) and Eq.~(\ref{reserror}) for the pure fixed-order and the resummation improved predictions, respectively. Although the resummed predictions usually have more conservative
handles over the perturbative uncertainties due to variation of multiple scales in the resummed
cross section, we still observe sizable reductions in the theoretical errors after invoking 
the resummation.

\begin{figure}[h!]
\begin{minipage}[b]{3.0in}
  \includegraphics[width=3.0in,angle=0]{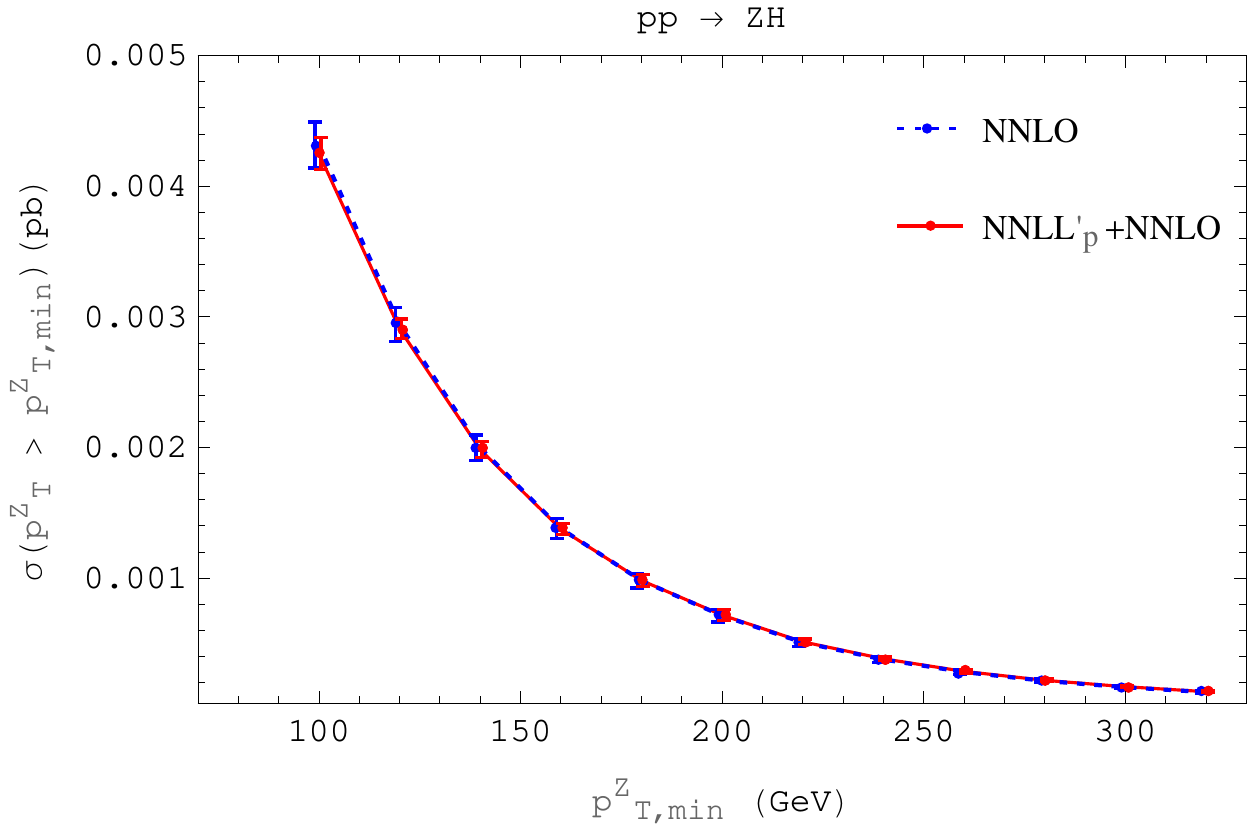}
\end{minipage}
\begin{minipage}[b]{3.0in}
  \includegraphics[width=3.0in,angle=0]{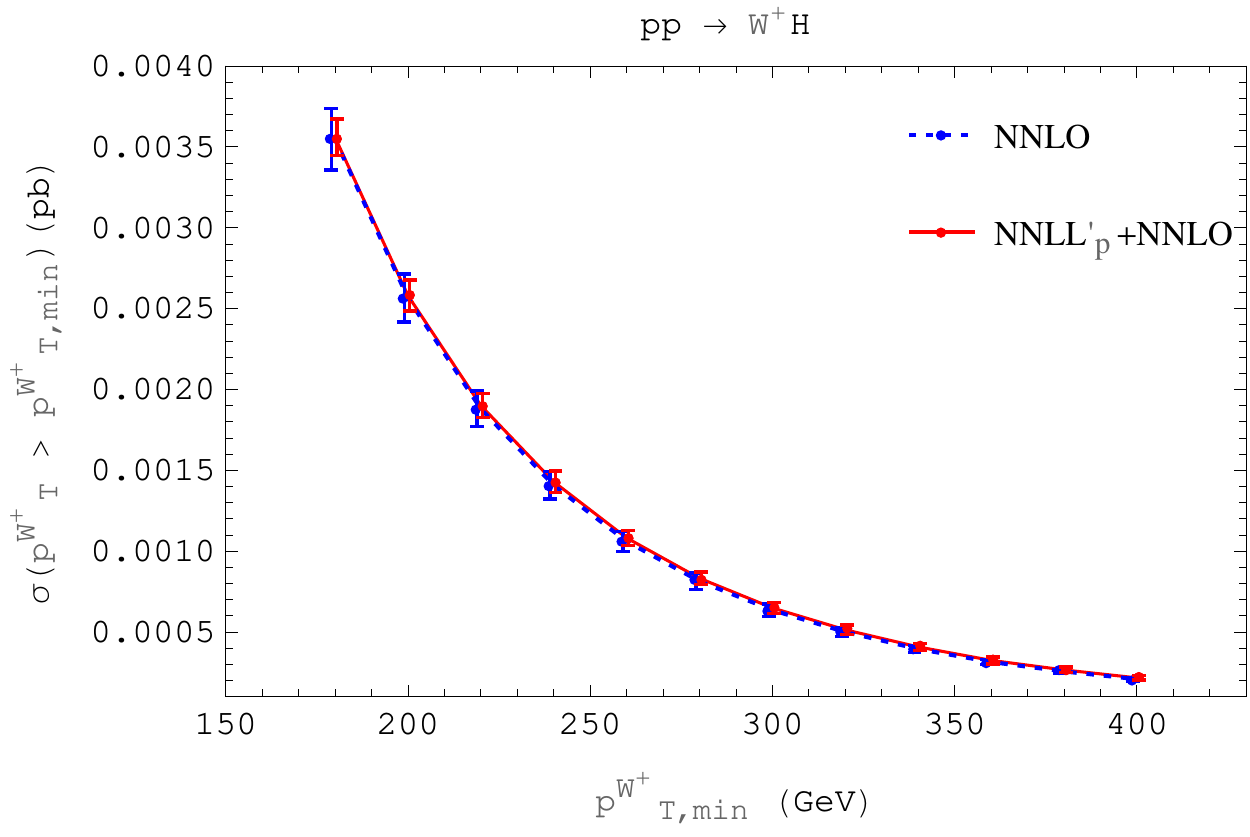}
\end{minipage}
\caption{The comparison between the NNLO cross sections with (red solid) and without (blue dotted) the ${\rm NNLL}'_p + \pi^2$ resummation for both $ZH$ (left panel) and $W^+H$ production at LHC$14$. The error bars reflect the scale uncertainties.}
\label{fig:nnlonnll}
\end{figure}

To see the reduction more clearly, we plot the NNLO and ${\rm NNLL}'_p+{\rm NNLO}$ predictions
together in fig.~\ref{fig:nnlonnll} for both $ZH$ and $W^+H$ productions. The pure fixed-order and the resummed calculations yield central values close to each other, yet the resummed cross section comes along with a reduced scale dependence. For the experimentally interesting region ($p^{Z,W}_{T,{\rm min}}>100,180$ GeV), the
scale dependence drops from $\pm 5(6)\%$ to $\pm3(4)\%$ for $ZH$ ($W^+H$) production 
at LHC$14$, {\it i.e.}  the theoretical error band shrinks by around $30\%$, and the veto efficiency for $ZH$ and $W^+H$ processes are roughly 47\% and 41\% respectively.

\section{Conclusions}\label{concl}
Recently, significant effort has been devoted to improved understanding of the Higgs
cross section at the LHC, either using conventional NNLO calculations~\cite{Boughezal:2013uia} or by utilizing resummation techniques~\cite{Banfi:2012yh,Banfi:2012jm,Becher:2012qa,Becher:2013xia,Stewart:2013faa,Liu:2012sz,Liu:2013hba,Boughezal:2013oha}.
In this work, we investigated the $VH$ associated production process at the LHC. On the experimental
side, a jet veto procedure demanding no jets with transverse momentum larger than $p_T^{veto}$
is used by both CMS and ATLAS to suppress the backgrounds in the boosted Higgs analysis.
However, theoretically the small value of $p_T^{veto} \ll m_V+m_H$ often destabilizes the 
perturbative expansion due to the existence of large logarithmic structure in the series. This makes
the perturbative prediction unreliable and results in large theoretical uncertainties on the cross section as the $p_T$ of the vector boson/Higgs increases. For $VH$ production, when the transverse momentum $p_{V}^T$ of the vector boson is
required to be larger than $100\sim 200\>{\rm GeV}$, the perturbative predictions suffer from roughly $\pm 6\%$ errors for LHC$14$ even at the NNLO. The limited power of the theoretical predictions restricts the accuracy that experimentalists can achieve.
Also the large K factors in the presence of jet veto when one goes from LO, NLO to NNLO leads to the 
concern that the missing higher orders corrections  may still be large and have sizable contributions to the cross section. 

Here we improved the theoretical predictions through the resummation the jet veto logarithms 
and a series of $\pi^2$ terms
up to 
${\rm NNLL}'_p$ accuracy within the SCET framework. We further matched the resummed
result onto the NNLO calculation to provide the full resummation improved cross section for $VH$
production at hadron colliders. The improved results reduce the theoretical uncertainties.
In the highly boosted regime, the scale uncertainty drops from $\pm 6\%$ to $\pm 4\%$ or below from NNLO to ${\rm NNLL}'_p+{\rm NNLO}$ for LHC$14$. Meanwhile resumming the jet veto logarithms greatly improves the convergence of the perturbative series in the region of interest experimentally. It can be seen from the fact that the central value remains virtually unchanged when one moves to higher orders in perturbative calculations, proving the reliability of the theoretical predictions.
After resummation, the jet veto efficiency is found to be $47\%$ and $41\%$ for $ZH$ and $W^+H$, respectively.

In this manuscript, we highlight the general features of the 
resummation improved predictions for $VH$ associated production, and show
the power of resummation in reducing scale uncertainties and providing reliable central values. 
For illustration purpose here, we have chosen simplified cuts from the current experimental analysis without loss of generality. 
The results can be extrapolated for experimental use. In the future, 
extensive numerical 
studies on $VH$ production with more practical cuts will be pursued in an upcoming paper,  based on the scheme in this work.

\section*{Acknowledgments}
We thank Frank Petriello for many helpful suggestions and carefully reading the manuscript.
We are grateful for discussions and kind supports from Stefan H\"{o}che on
cross-checking FEWZ. We thank Jike Wang for explaining current experimental status.
X.L. would like to thank Yin Cui for all the supports when this work was carried out. Y.L. was supported by the US Department of Energy
under contract DE--AC02--76SF00515. X.L. was supported by DE-AC02-06CH11357 and the grants  DE-FG02-95ER40896 and DE-FG02-08ER4153. Y.L would like to thank for the hospitality of Northwestern University during the completion of this work. This research used resources of the National Energy Research Scientific Computing Center, which is supported by the Office of Science of the U.S. Department of Energy under Contract No. DE-AC02-05CH11231.

\appendix

\section{modification of FEWZ \label{sec:fewzh}}

Here we give a brief introduction of the master formula we use to modify FEWZ for the Higgsstrahlung process. The squared matrix element of the DY process in FEWZ can be schematically written as,
\bea
\overline{|\mathcal{M}_{V}|^2} &\sim& Q_{\mu\nu}L^{\mu\nu},
\eea
where $Q_{\mu\nu}$ is the square of the quark current and includes all QCD corrections, and $L^{\mu\nu}$ is the square of the lepton current. As long as no observables related to asymmetry in the lepton phase space, we can neglect the axial part of the current, hence
\bea
L^{\mu\nu} &\sim& -g^{\mu\nu} l_1\cdot l_2 + l_1^\mu ~l_2^\nu + l_1^\nu ~l_2^\mu.
\eea
When integrated over the lepton phase space inclusively, we find that, 
\bea
\langle L^{\mu\nu} \rangle &\sim& \frac{q^2}{3}  \left(-g^{\mu\nu} + \frac{q^\mu q^\nu}{q^2} \right),
\label{eq:fewzhkey1}
\eea
with $l_{1,2}$ as the momenta of the two leptons and $q=l_1+l_2$ as the momentum of the vector boson. In original FEWZ, individual components of $Q_{\mu\nu}$ are not separately available since they only appear in the product of two currents in the matrix element.
For the DY-like part of the $VH$ production, it can be thought as a gauge boson $V^*$ first produced via the DY process, and subsequently decaying to $V$ and $H$, where $V$ further decays to two leptons. Because of gauge invariance, we can write the new squared matrix element for Higgsstrahlung process as,
\bea
\overline{|\mathcal{M}_{VH}|^2} &\sim& Q_{\mu\nu}L'^{\mu\nu},
\eea
where $L'^{\mu\nu}$ denotes the square of the new lepton current. If we integrate inclusively over the $V$ and $H$ phase space, the new lepton current becomes,
\bea
\langle L'^{\mu\nu}\rangle &\sim& \frac{q'^2}{3}  \left(-g^{\mu\nu} + \frac{q'^\mu q'^\nu}{q'^2} \right) \nn\\
&\rightarrow& \frac{q'^2}{3}  \left\{-g^{\mu\nu} \left(\frac{2}{3}+\frac{(q\cdot q')^2}{3\, q^2\,q'^2}\right) + \frac{q^\mu q^\nu}{q^2} \left(-\frac{1}{3}+\frac{4}{3}\frac{(q\cdot q')^2}{q^2\,q'^2}\right) \right\},
\eea
in which $q'$ and $q$ are the momenta of $V$ and $V^*$ respectively, and $q\cdot q'=(q^2+q'^2-m_H^2)/2$. The second piece drops out when multiplied with the quark current, and therefore we arrive at,
\bea
\label{eq:fewhmodification1}
Q_{\mu\nu}\langle L'^{\mu\nu} \rangle = \frac{q'^2}{q^2} Q_{\mu\nu}\langle L^{\mu\nu} \rangle \left(\frac{2}{3}+\frac{(q\cdot q')^2}{3\, q^2\,q'^2}\right).
\eea
However, Eq.~(\ref{eq:fewhmodification1}) does not apply to the case where acceptance cut is placed on leptons, the vector boson or the Higgs. We can use 
\bea
\langle L^{\mu\nu}L^{\rho\sigma} \rangle &=& \frac{q^4}{4} \left\{ \frac{2}{5} \left( -g^{\mu\nu}+\frac{q^\mu q^\nu}{q^2} \right) \left( -g^{\rho\sigma}+\frac{q^\rho q^\sigma}{q^2} \right) + \right. \nn\\
&& \left. \frac{1}{15} \left[ \left( -g^{\mu\rho}+\frac{q^\mu q^\rho}{q^2} \right) \left( -g^{\nu\sigma}+\frac{q^\nu q^\sigma}{q^2} \right) + \left( -g^{\mu\sigma}+\frac{q^\mu q^\sigma}{q^2} \right) \left( -g^{\nu\rho}+\frac{q^\nu q^\rho}{q^2} \right) \right] \right\}, \nn\\
\label{eq:fewzhkey2}
\eea
in combination with Eq.~(\ref{eq:fewzhkey1}) and obtain
\bea
\label{eq:fewhmodification2}
Q_{\mu\nu} L'^{\mu\nu} = 30\, Q_{\mu\nu} \frac{\langle L^{\mu\nu} L^{\rho\sigma} \rangle}{q^4} L'_{\rho\sigma} - 9\, Q_{\mu\nu} \frac{ \langle L^{\mu\nu} \rangle}{q^2} \left(-g^{\rho\sigma} + \frac{q^\rho q^\sigma}{q^2} \right) L'_{\rho\sigma}.
\eea
The average is performed by integrating over the original lepton phase space inclusively, and arbitrary cuts can be applied on the new vector boson and Higgs, as well the final leptonic final states, as long as no asymmetry related observables are measured. It can be seen that Eq.~(\ref{eq:fewhmodification1}) is easily recovered by substituting the integrated form of $L'_{\rho\sigma}$ into Eq.~(\ref{eq:fewhmodification2}). Because only the product $Q_{\mu\nu} L^{\mu\nu}$ appears in Eq.~(\ref{eq:fewhmodification2}), FEWZ can be modified in a relatively straightforward way.

\section{Fixed-order matrix elements}
In this appendix, we list all the ingredients needed for a
 $\text{NNLL}^\prime$ resummation for $pp \to VH$ production in $0$-jet bin.
We start with the fixed order matrix elements to ${\cal O}(\alpha_s^2)$

\subsection{Hard Function}

The spin and color averaged LO matrix elements squared for processes $q{\bar q} \to ZH \to {\bar l}lH$ 
and $q{\bar q} \to WH \to \nu l H$ are proportional to that of LO DY processes,
\bea
\overline{|\mathcal{M}_{Z}|^2}&=&\frac{1}{12} (4 \pi \alpha)^2 \left(  \frac{ 4 (|g^+_{qqZ} g^+_{llZ}|^2+|g^-_{qqZ} g^-_{llZ}|^2) s_{q\bar{l}} s_{\bar{q}l} + 4 (|g^+_{qqZ}g^-_{llZ}|^2+|g^-_{qqZ} g^+_{llZ}|^2) s_{ql} s_{\bar{q}\bar{l}} } {(s_{l\bar{l}}-M_Z^2)^2+ M_Z^2 \Gamma_Z^2} \right) \,, \nn\\
\overline{|\mathcal{M}_{W^-}|^2}&=&\frac{1}{12} \frac{(4 \pi \alpha)^2}{s_w^4} \frac{s_{q\bar{\nu}} s_{\bar{q}l}}{(s_{l\bar{\nu}}-M_W^2)^2+ M_W^2 \Gamma_W^2} \,, \nn\\
\overline{|\mathcal{M}_{W^+}|^2}&=&\frac{1}{12} \frac{(4 \pi \alpha)^2}{s_w^4} \frac{s_{q\bar{l}} s_{\bar{q}\nu}}{(s_{\nu\bar{l}}-M_W^2)^2+ M_W^2 \Gamma_W^2} \,, 
\eea
with $s_{ij}= 2\,p_i\cdot p_j$\,, $g^+_{ffZ}=-Q_f s_w/c_w$ and $g^-_{ffZ}=(I^3_{W,f}-Q_f s_w^2)/s_w/c_w$.
Therefore 
\bea
\overline{|\mathcal{M}_{ZH}|^2}&=&\frac{4 \pi \alpha M_Z^2}{s_w^2 c_w^2} \frac{1}{(s_{q\bar{q}}-M_Z^2)^2+ M_Z^2 \Gamma_Z^2}  \overline{|\mathcal{M}_{Z}|^2}  \,, \nn\\
\overline{|\mathcal{M}_{W^\pm H}|^2}&=&\frac{4 \pi \alpha M_W^2}{s_w^2} \frac{1}{(s_{q\bar{q}}-M_W^2)^2+ M_W^2 \Gamma_W^2}  \overline{|\mathcal{M}_{W^\pm}|^2} \,,
\eea
where $M_Z$ and $M_W$ are the masses for $Z$ and $W$ bosons, respectively 
and $\Gamma_Z$ and $\Gamma_W$ are the widths. 

The $\text{NNLO}$ hard function for Drell-Yan has been known for a while and can be found for instance
in~\cite{Becher:2007ty}, which gives 
 \bea
H_{q\bar{q}V}(Q^2,\mu) &=& 1+ 
\frac{\alpha_s}{4\pi}  C_F \left(
-2L^2 - 6L
-16+\frac{7\pi^2}{3}
\right)\, \nn\\
&+&
\left( \frac{\alpha_s}{4\pi}\right)^2 \left[ 2 C_F \left(
C_F H_F + C_A H_A + T_F n_f H_f
\right) + C_F^2  G_F \right] \,,
\eea
with
\bea
&&H_F = \frac{L^4}{2}  + 3 L^3 
+\left(\frac{25}{2}-\frac{19\pi^2}{6} \right)L^2
+\left(\frac{45}{2}-\frac{15\pi^2}{2}-24\zeta_3\right)L
+\frac{255}{8} -9\pi^2 +\frac{157\pi^4}{360}-30\zeta_3\,, \nn \\
&&H_A = -\frac{11}{9}L^3 \,
-\left(\frac{233}{18}-\frac{\pi^2}{3} \right)L^2 \,
- \left(\frac{2545}{54}-\frac{22\pi^2}{9}-26\zeta_3 \right)L
-\frac{51157}{648} 
+ \frac{1061\pi^2}{108}
 -\frac{4\pi^4}{45}
 + \frac{313}{9}\zeta_3\,,\nn\\
&& H_f =  \frac{4}{9}L^3 +\frac{38}{9}L^2 \,
 +\left( \frac{418}{27} -\frac{8\pi^2}{9}\right)L\,
 + \frac{4085}{162} - \frac{91\pi^2}{27}+\frac{4}{9}\zeta_3 \,, \nn\\
&& G_F = L^4+6 L^3+\left( 25+\frac{5 \pi^2}{3}\right) L^2+\left(48+5\pi^2\right) L+64-\frac{29 \pi^2}{3}+\frac{49 \pi^4}{36} .
\eea
Here we have abbreviated $L \equiv \log(\mu^2/Q^2)$.

\subsection{Soft Function}
The full two-loop soft function can be extracted from Ref.~\cite{Stewart:2013faa} by suitably 
replacing the color factors, and is found to have the form
\bea \label{eq:softFO}
 S_{q{\bar q}} (p_T^{veto}, R, \mu, \nu) &=&
1 + \frac{\alpha_s(\mu)}{4\pi} C_F \Bigl[ 2\Gamma_0 L_S^\mu \bigl( L_S^\mu - 2 L_S^\nu) - \frac{\pi^2}{3} \Bigr]
\nn \\ 
&+& 
\frac{\alpha_s^2(\mu)}{(4\pi)^2}\, \biggl\{
\frac{1}{2}C_F^2 \Bigl[ 2\Gamma_0  L_S^\mu \bigl( L_S^\mu - 2 L_S^\nu) - \frac{\pi^2}{3}  \Bigr]^2
+ 2\beta_0 C_F L_S^\mu \Bigl[2\Gamma_0  L_S^\mu\Bigl(\frac{1}{3} L_S^\mu - L_S^\nu\Bigr) - \frac{\pi^2}{3}  \Bigr]
\nn \\
 &+& 
2\Gamma_1 C_F L_S^\mu (L_S^\mu - 2 L_S^\nu)
+ \gamma_{S\,1}^q L_S^\mu + \gamma_{\nu\, 1}^q(R)\, L_S^\nu + s_2(R) \biggr\}
\,,
\eea
with 
$L_S^\mu \equiv \log \frac{\mu}{p_T^{veto}}$, and $L_S^\nu \equiv \log \frac{\nu}{p_T^{veto}}$.

The dependence on the jet algorithm starts to enter at two loops through the two-loop $\nu$ anomalous dimension, $\gamma_{\nu\, 1}^g(R)$, which determines the coefficient of the single logarithm of $\ln(\nu/p_T^{veto})$, as well as the non-logarithmic two-loop soft constant, $s_2(R)$. For the coefficients of the soft non-cusp anomalous dimensions we find
\bea\label{eq:gammaS1munu}
\gamma_{S\, 0}^q &= &0 \,, \nn \\
\gamma_{S\, 1}^q
&= & 8 C_F \biggl[ \Bigl(\frac{52}{9} - 4 (1+\pi^2) \ln 2 + 11 \zeta_3\Bigr) C_A
+ \Bigl(\frac{2}{9} + \frac{7 \pi^2}{12} - \frac{20}{3} \ln 2 \Bigr) \beta_0 \biggr]
\,, \nn \\
\gamma_{\nu\, 0}^q(R) &=& 0 \,, \nn \\
\gamma_{\nu\, 1}^q(R)
&= & -16 C_F \biggl[\Bigl(\frac{17}{9} - (1 + \pi^2)\ln2 + \zeta_3 \Bigl) C_A
+ \Bigl(\frac{4}{9} +\frac{\pi^2}{12} - \frac{5}{3}\ln2 \Bigl) \beta_0 \biggr] + C_2(R)
\,.
\eea
Here, $C_2(R)$ is the clustering correction due to the jet algorithm:
\bea\label{eq:C2value}
C_2(R) &=&
 2 C_F \Bigl[\Bigl(1 - \frac{8\pi^2}{3}\Bigr) C_A + \Bigl(\frac{23}{3} - 8\ln2\Bigr)\beta_0\Bigl] \ln R^2
\nn\\ & &
+ 15.62 C_F C_A - 9.17 C_F \beta_0
\,,
\eea
up to ${\cal O} (R^2)$ corrections whose explicit form can be lifted from 
Ref.~\cite{Becher:2013xia}.

The two-loop soft function constant $s_2(R)$, which is not determined from RGE constraints, is
\bea \label{eq:s2R}
s_2 (R) &=&
C_F \biggl[ \Bigl(\frac{19}{3} - 10 \ln 2 + 8 \zeta_3 \Bigr) C_A
+ \Bigl(-\frac{163}{9} + \frac{58}{3}\ln2 + 8 \ln^2 2 \Bigr) \beta_0 \biggr] \ln R^2
\nn\\ &&
-18.68 C_F C_A - 3.25 C_F \beta_0 + s_2^{{\rm R}_{\rm sub}}(R)
\,,
\eea
where $s_2^{{\rm R}_{\rm sub}}(R) \sim R^2$. 

\subsection{Beam Function}
When matching onto pdfs, the $\text{NNLO}$ beam function has the form
\bea
B_q(x,\mu_B,\nu_B) = \sum_j \int_x^1 \frac{\mathrm{d}z}{z} \,
{\cal I}_{qj} (R,z,\mu_B,\nu_B)\,
 f_j\left(\frac{x}{z},\mu_B \right)\,,
\eea
where the matching kernels ${\cal I}_{qj}$ read
\bea
{\cal I}_{qj} = \delta(1-z)\delta_{qj} + \frac{\alpha_s}{4\pi}{\cal I}_{qj}^{(1)}\,
+ \left(\frac{\alpha_s}{4\pi}\right)^2 {\cal I}_{qj}^{(2)} \,,
\eea
with
\bea
&&{\cal I}_{q_iq_i}^{(1)} = 2C_F  \left( 4L_{\mu}\,L_\nu \delta(1-z)\,
 - 2L_{\mu}\, p_{q_iq_i}^{(0)}(z) + I_{q_iq_i}^{(1)}(z) \right)\,, \nn \\
&&{\cal I}_{qg}^{(1)} = 2T_F \left( -2 L_\mu \,p^{(0)}_{qg}(z)+I_{qg}^{(1)}(z)   \right) \,,
\eea
given that $L_\mu \equiv \log \frac{\mu}{p_T^{veto}}$ and
$L_\nu \equiv \log \frac{\nu}{Q}$.

The form of the two-loop matching coefficient ${\cal I}^{(2)}$ can be obtained
by expanding the $\text{NNLL}^\prime$ resummed beam function and matching
onto PDFs, which results in
\bea
{\cal I}_{q_iq_i}^{(2)} (z) &=& 
L_\mu^2 L_\nu^2\,  2\Gamma_0^2 C_F^2\, \delta(1-z) \,
+ L_\mu^2 L_\nu \, 
\left(2\Gamma_0C_F\beta_0\delta(1-z) - 8 C^2_F\Gamma_0\, p_{q_iq_i}^{(0)}(z) \right) \nn \\
&+& \, L_\mu L_\nu\,
\left( 4\Gamma_0 C_F^2 I^{(1)}_{q_iq_i}(z) + 2\Gamma_1C_F\, \delta(1-z) \right) \nn \\
&+& \, L_\mu^2 \, \left( -4C_F\beta_0 p_{q_iq_i}^{(0)}(z)\,
 + 8 C_F^2\,p^{(0)}_{q_iq_i}\otimes p_{q_iq_i}^{(0)}(z)
 + 8 C_FT_F\, p^{(0)}_{q_ig}\otimes p_{gq_i}^{(0)}(z) \right) \nn \\
 &+&  \, L_\mu \,
 \left\{ 
 4C_F\beta_0 I_{q_iq_i}^{(1)}(z) + \gamma^q_{B,1}\delta(1-z) \,
 -8 {\bar p}_{q_iq_i}^{(1)}(z) \,
  -8C_F^2\, I_{q_iq_i}^{(1)} \otimes p_{q_iq_i}^{(0)} \,
  -8 C_F T_F\, I_{q_ig}^{(1)} \otimes p_{gq_i}^{(0)}
 \right\} \nn \\
 &+&  \, L_\nu\, \left( -\frac{1}{2} \gamma^q_{\nu,1}\delta(1-z) \right) 
 + I_{q_iq_i}^{(2)}(R,z)
 \,,
\eea
\bea
{\cal I}_{q_ig}^{(2)} (z) &=& 
L_\mu^2 L_\nu \left( -8\Gamma_0 C_FT_F\, p_{q_ig}^{(0)}(z) \right) \,
+ L_\mu L_\nu \left( 4\Gamma_0 C_F T_F\, I_{q_ig}^{(1)}(z) \right) \, \nn \\
&+& L_\mu^2 \, 
\left\{ 
-\left(4\beta_0 + 2\gamma^q_{B,0} \right)T_F\, p_{q_ig}^{(0)}(z)
+ 8 C_F T_F \, p_{q_iq_i}^{(0)} \otimes p_{q_ig}^{(0)}(z)
+ 8 C_A T_F\, p_{q_ig}^{(0)} \otimes {\bar p}_{gg}^{(0)}(z) 
\right\} \nn \\
&+& L_\mu \, 
\left\{
(4\beta_0 + 2\gamma^q_{B,0})T_F\, I_{q_ig}^{(1)} (z)
-8 p_{q_ig}^{(1)}(z) \,
-8C_FT_F\, I_{q_iq_i}^{(1)} \otimes p_{q_ig}^{(0)}(z)\,
-8 C_A T_F\, I_{q_ig}^{(1)} \otimes {\bar p}_{gg}^{(0)} (z)
\right\} \nn \\
 &+& I_{q_ig}^{(2)}(R,z)
\,,
\eea
and
\bea\label{eq:Iqiqj}
{\cal I}_{q_i q_j}^{(2)}(z)& =& \,
L_\mu^2 \, \left( 8C_FT_F\,p_{q_ig}^{(0)} \otimes p_{gq_j}^{(0)} (z)  \right)\,
+ L_\mu \, \left( -8  p_{q_iq_j}^{(1)}(z) -8C_FT_F\, I_{q_ig}^{(1)} \otimes p_{gq_j}^{(0)}(z) \right) \nn \\
&&+ I_{q_iq_j}^{(2)}(R,z)
\,,
\eea
where ${\bar p}_{kj}^{(i)}(z)$ is the $i$-th order 
full splitting function including properly the $\delta_{kj}\delta(1-z)$ while $p_{kj}^{(i)}(z)$ does not. 
In both cases, an overall color factor has been extracted from the splitting kernel. In Eq.~(\ref{eq:Iqiqj}),
$q_j$
in the subscript
stands for either quarks with different flavors from $q_i$ or any possible anti-quarks.

The non-logarithmic terms $I_{ij}^{(2)}(R,z)$ can not be determined by the expansion and have to be
computed explicitly, and are not yet known. However, for small $R$, the dominant $\log R$ piece can be 
calculated in a very simple way, as explained in the text, which leads to
\bea
&&I_{q_iq_i,\log R}^{(2)}  = 
\frac{2}{9}C_F\left( 
\left(-12 \pi ^2+131-132 \log (2)\right) C_A
+  (24 \log (2)-23) n_f T_F
\right)
\log(R^2) \, p_{q_iq_i}^{(0)}(z)\,,
\nn \\
&&I_{q_ig,\log R}^{(2)}  =\,
4 C_FT_F \left(-\frac{\pi ^2}{3}+3-3 \log (2)\right)
  \log(R^2)
 p_{q_i,g}^{(0)}(z)\,.
\eea
The contribution from the remaining piece can be obtained by fitting
with the fixed order $\text{NNLO}$ QCD calculation.

\section{RG running}
All the functions in the previous section have to be evolved 
from their natural scales to a common scale $\mu$
to evaluate the cross section. Other than the conventional RG evolution, 
\bea
\mu \frac{\mathrm{d} F}{\mathrm{d} \mu} = \Gamma_F^\mu (\mu) F(\mu) \,,
\eea
due to 
the existence of the rapidity divergence in SCET resulted from the multipole expansion,
another rapidity evolution
\bea
\nu \frac{\mathrm{d} F_{B,S}}{\mathrm{d} \nu} = \Gamma^\nu_{B,S} (\nu) F_{B,S}(\nu) \,,
\eea
 for the soft
and the beam functions is needed to resum a series of large rapidity logs.
The general solution to these RG equations can be formally written as
\bea
F(\mu, \nu) = U(\mu, \nu; \mu_0, \nu_0) F(\mu_0, \nu_0)\,,
\eea
where the natural scales $(\mu_0, \nu_0)$ for each function are determined by demanding that no large
logs exist in the fixed order matrix elements. 
The evolution of the hard Wilson coefficient $C_H^{q{\bar qV}}$ which 
is related to the hard function by $H = C C^\dagger $, is given by
\bea
U_{C_H}(\mu,\mu_H) = \exp\left( 
2C_F S(\mu,\mu_H) - C_F A_\Gamma \log \frac{-Q^2 - i0^+}{\mu^2_H}\,
-A_H(\mu, \mu_H)
\right)\,,
\eea
 where a natural choice of $\mu_H$ will be $\mu^2_H = -Q^2 - i0^+$ to stabilize the
 fixed order expansion of the hard function.  By doing so, a towers of $\pi^2$ terms
 will also be resummed. Therefore the evolution of the hard function is given by
 \bea
 U_H &= &U_H^{\rm log} \times 
 \exp\left(2\Re e\left[2C_F S(\mu_H, - \mu_H) -C_F A_\Gamma(\mu_H, - \mu_H)
 \log \frac{Q^2}{\mu_H^2} - A_H(\mu_H, -\mu_H) 
  \right] \right) \nn \\
  &=& U_H^{\rm log}(\mu,\mu_H) \,
  \exp\left( \frac{\pi\alpha_s(\mu_H) C_F}{2}
  \left[1 + \frac{1}{4\pi}\left( \frac{\Gamma_1}{\Gamma_0}
  - \frac{\gamma_H^0}{\Gamma_0}  \frac{\beta_0}{C_F}\,
  -\beta_0 \log \frac{Q^2}{\mu_H^2}
    \right)\alpha_s(\mu_H) \right]
   \right) \,.
 \eea
Here $U_H^{\rm log} = U_C U_C^\dagger$ is the normal evolution for the global log resummation.

The running of the beam function is found to be
\bea
U_{B,a}(\mu,\nu;\mu_B,\nu_B) &=& \,
\exp\left(-C_F\, A_\Gamma\left(\mu,p_T^{veto} \right)\log\frac{\nu^2}{\nu^2_B} \right)\,
\exp\left(-C_F\, A_\Gamma(\mu,\mu_B ) \log \frac{\nu_B^2}{\w_a^2}\,
-A_{B_a}(\mu,\mu_B)
 \right) \,, \nn \\
 && \times \exp\left(\,
 -\frac{1}{2}\gamma^q_\nu\left[\alpha_s(p_T^{veto}),R\right] \log\frac{\nu}{\nu_B}
 \right) \,.
\eea 
The central value of $(\mu_B,\nu_B)$ will be chosen as $\left(p_T^{veto}, \w_a\right)$.

For the soft function, we have
\bea
U_S(\mu, \nu; \mu_S, \nu_B) &=& \,
\exp\left( 2C_F\, A_\Gamma(\mu, p_T^{veto})\log \frac{\nu^2}{\nu_S^2} \right) \,
\exp\left( -4C_F S(\mu,\mu_S) - A_S(\mu,\mu_S)\right)\, \nn \\
&&\times \exp\left( 2C_F \, A_\Gamma(\mu,\mu_S)\log \frac{\nu_S^2}{\mu_S^2}   \right)
\exp\left(\gamma^q_{\nu}\left[\alpha_s(p_T^{veto}),R\right] \log\frac{\nu}{\nu_S}\right)
\,.
\eea
The natural scale $(\mu_S, \nu_S)$ for the soft sector is $(p_T^{veto}, p_T^{veto})$.

In the equations above, the expansion of these quantities in $\alpha_s$ up to terms needed for NNLL resummation are given by
\bea
S(\mu_f,\mu_i) &=& \frac{\Gamma_0}{4\beta_0^2}\Bigg \{ \frac{4\pi}{\alpha_s(\mu_i)} \Big ( 1-\frac{1}{r}-\ln r\Big ) + \Big ( \frac{\Gamma_1}{\Gamma_0}-\frac{\beta_1}{\beta_0}\Big )(1-r+\ln r )+\frac{\beta_1}{2\beta_0}\ln^2r\nn \\
&+& \frac{\alpha_s(\mu_i)}{4\pi}\Bigg [\Big ( \frac{\beta_1\Gamma_1}{\beta_0\Gamma_0} - \frac{\beta_2}{\beta_0} \Big )(1-r+r\ln r) +\Big ( \frac{\beta_1^2}{\beta_0^2}-\frac{\beta_2}{\beta_0}\Big )(1-r)\ln r \nn \\
&-&\Big(\frac{\beta_1^2}{\beta_0^2}-\frac{\beta_2}{\beta_0}-\frac{\beta_1 \Gamma_1}{\beta_0\Gamma_0}+\frac{\Gamma_2}{\Gamma_0}\Big)\frac{(1-r)^2}{2}\Bigg ]\Bigg \} \,,
\eea
with $r = \alpha_s(\mu_f)/\alpha_s(\mu_i)$, 
and
\bea
\label{Aevo}
A_\Gamma(\mu_f,\mu_i)&=& \frac{\Gamma_0}{2\beta_0}\left\{
\log r + \,
\frac{\alpha_s(\mu_i)}{4\pi}\left(\frac{\Gamma_1}{\Gamma_0}-\frac{\beta_1}{\beta_0} \right)\,
(r-1) \,
\right.\nn \\
&&\left.\,
+ \frac{\alpha_s^2(\mu_i)}{16\pi^2}\,
\left[\frac{\Gamma_2}{\Gamma_0}-\frac{\beta_2}{\beta_0}
-\frac{\beta_1}{\beta_0}\,
\left(\frac{\Gamma_1}{\Gamma_0}-\frac{\beta_1}{\beta_0} \right)
\right]\frac{r^2-1}{2}
\right\}.
\eea
Also $A_H$, $A_B$ and $A_S$ are needed to the $\alpha_s$ order, which 
can be obtained by substituting the $\Gamma_0$ and $\Gamma_1$ in
$A_\Gamma$ with $\gamma_{i,0}$ and $\gamma_{i,1}$ for each function
and truncating out the $\alpha_s^2$ terms. 

\section{Input ingredients}
Here we group all the parameters and equations including splitting functions and convolutions
which are used in $\text{NNLL}^\prime$ resummation

The $0$-th order modified splitting kernels needed in the beam function are
\bea
&&{p}^{(0)}_{gg}(z) = \frac{2z}{(1-z)_+}+2z(1-z)+2\frac{1-z}{z}\,, \nn\\
&&{p}^{(0)}_{qq}(z) = \frac{1+z^2}{(1-z)_+}  \,, \nn \\
&&p^{(0)}_{gq}(z) = \frac{1+(1-z)^2}{z} \,, \nn \\
&&p^{(0)}_{qg}(z) = 1 -2z +2z^2\,,
\eea
and
\bea
&&{\bar p}_{q_iq_j}^{(0)}(z) = \frac{3}{2}\delta_{ij} \delta(1-z) + p_{qq}^{(0)}(z)\,,\nn \\
&&{\bar p}_{gg}^{(0)}(z) = \frac{\beta_0}{2C_A}\delta(1-z) + p_{gg}^{(0)}(z) \,. 
\eea
The $1$-th order splitting functions give~\cite{Ellis:1996nn}
\bea \label{pqiqj2}
&&p_{q_iq_j}^{(1)}=\delta_{ij} p_{qq}^{V,(1)} + p_{qq}^{S,(1)} \,,\nn\\
&&p_{q_i\bar{q}_j}^{(1)}=\delta_{ij} p_{q\bar{q}}^{V,(1)} + p_{q\bar{q}}^{S,(1)}\,,
\eea
and
\bea \label{pqq2}
&&p_{qq}^{V,(1)}=
  C_F^2 \Big\{ -\big[2 \ln x \ln(1-x)+\frac{3}{2} \ln x  \big] p_{qq}^{(0)}(x) 
\nonumber \\ &&
 -\left(\frac{3}{2}+\frac{7}{2} x\right)\ln x
      -\frac{1}{2} (1+x) \ln^2 x -5 (1-x)\Big\}
\nn \\ &&
      +C_F C_A \Big\{ \left[\frac{1}{2} \ln^2 x
 +\frac{11}{6} \ln x+\frac{67}{18}-\frac{\pi^2}{6} \right] p^{(0)}_{qq}(x)
      +(1+x) \ln x+\frac{20}{3} (1-x)\Big\} 
\nn \\ &&
      +
n_f C_F T_F \Big\{-\left[\frac{2}{3} \ln x+\frac{10}{9}\right] p^{(0)}_{qq}(x) 
 -\frac{4}{3}  (1-x))\Big\} \,, \nn\\
&&p_{q\bar{q}}^{V,(1)} = C_F \left(C_F-\frac{C_A}{2}\right) 
\Big\{2 p_{qq}^{(0)}(-x) S_2(x)+2 (1+x) \ln x+4 (1-x) \Big\}   \; ,
\eea
\bea \label{pff2}
&&p_{qq}^{S,(1)} = p_{q\bar{q}}^{S,(1)} \nn\\
&&= C_F T_F \Big\{ \frac{20}{9x} -2 +6 x-\frac{56}{9} x^2
+\left(1+5 x+\frac{8}{3} x^2\right) \ln x - (1+x) \ln^2 x \Big\} \; ,
\eea
\bea \label{pqg2}
&&p_{qg}^{(1)}=\frac{C_F T_F}{2}
       \Big\{4-9 x-(1-4 x) \ln x-(1-2 x) \ln^2x +4 \ln(1-x)        
\nn \\ &&
      +\left[2 \ln^2\left(\frac{1-x}{x}\right)-4 \ln\left(\frac{1-x}{x}\right)
      -\frac{2}{3} \pi^2 +10 \right] p_{qg}^{(0)}(x)\Big\} \nn \\ &&      
      +\frac{C_A T_F}{2}
       \Big\{\frac{182}{9}+\frac{14}{9} x+\frac{40}{9 x}
  +\left(\frac{136}{3} x-\frac{38}{3}\right) \ln x-4 \ln(1-x) 
      -(2+8 x) \ln^2x \nn \\ && 
       +\left[-\ln^2x+\frac{44}{3} \ln x-2 \ln^2(1-x)+4 \ln(1-x)
+\frac{\pi^2}{3}
 -\frac{218}{9}\right] p^{(0)} _{qg}(x)
 \nn \\ &&
  +2 p_{qg}^{(0)}(-x) S_2(x) \Big\} \; ,
\eea
where the function $S_2(x)$ is defined as
\bea
S_2(x) &=& \int_{\frac{x}{1+x}}^{\frac{1}{1+x}} \frac{dz}{z} 
\ln \big(\frac{1-z}{z}\big)  = \frac{\ln^2 x}{2}-\ln x \ln(1+x) + 
{\rm Li}_2 \left(\frac{x}{1+x}\right)-{\rm Li}_2 \left(\frac{1}{1+x}\right) \nn\\
&=&  \frac{\ln^2 x}{2}- \frac{\pi^2}{6} - 2 \ln x \ln(1+x) -2\, {\rm Li}_2 (-x) 
\;.
\eea
To extend to the limit $x=1$, we need to make the substitution
\bea \label{plusify}
\frac{1}{1-x} \rightarrow \frac{1}{[1-x]_+}  \,.
\eea
and add the end-point contributions:
\bea
{\bar p}_{q_iq_i}^{(1)}(x) = p_{q_iq_i}^{(1)}(x)
&+&\Bigg[ 
C_F^2 \Big\{\frac{3}{8}-\frac{\pi^2}{2}+6 \zeta_3 \Big\}
+C_F C_A \Big\{\frac{17}{24}+\frac{11\pi^2}{18}-3 \zeta_3 \Big\}
\nonumber \\ && 
-n_f C_F T_F \Big\{\frac{1}{6}+\frac{2 \pi^2}{9}\Big\}\Bigg]\delta(1-x) \,.
\eea

The convolutions needed for evaluating the two-loop beam function are
\bea
&&p_{qq}^{(0)}\otimes p_{qq}^{(0)}(z) = -2 (1-z)+3 (1+z) \log (z)
-4(1+z)\log(1-z) \, \nn \\
 &&\hspace{8.3ex} \, + \,
8\left(\frac{\log(1-z)}{1-z} \right)_+ - \frac{2\pi^2}{3}\delta(1-z) - 4 \frac{\log(z)}{1-z}\,,\nn \\
&&p^{(0)}_{qg}\otimes p^{(0)}_{gq} (z)= -\frac{4 z^2}{3}-z+\frac{4}{3 z}+2 (z+1) \log (z)+1\,, \nn \\
&&p_{qq}^{(0)}\otimes p^{(0)}_{qg}(z) =-3 z^2+\left(-4 z^2+2 z-1\right) \log (z)+5 z-2 + 2p^{(0)}_{qg}(z)\log(1-z)\,, \nn \\
&& p^{(0)}_{qg}\otimes {\bar p}^{(0)}_{gg}(z)=-\frac{31 z^2}{3}+8 z+\frac{4}{3 z}+(8 z+2) \log (z)\,
+2p^{(0)}_{qg}(z)\log(1-z)+1 + \frac{\beta_0}{2C_A} p^{(0)}_{qg}(z)\,,\nn\\
&&I^{(1)}_{qq}\otimes p^{(0)}_{qq}(z)= -(1-z) (\log (z)+2) + 2I_{qq}^{(1)}(z)\log(1-z)\,,\nn\\
&&I^{(1)}_{qg}\otimes p^{(0)}_{gq}(z)=\frac{2}{3} \left(2 z^2+\frac{1}{z}-3 z \log (z)-3\right)\,,\nn\\
&& I^{(1)}_{qg}\otimes {\bar p}^{(0)}_{gg}(z) = \frac{2}{3} \left(17 z^2-15 z+\frac{1}{z}-12 z \log (z)-3\right)
+2I_{qg}^{(1)}(z) \log(1-z) + \frac{\beta_0}{2C_A} I_{qg}^{(1)}(z) 
\,, \nn\\
&&I^{(1)}_{qq}\otimes p^{(0)}_{qg}(z) = z^2+z-(2 z+1) \log (z)-2\,, 
\eea
with
\bea
&&I_{qq}^{(1)}(z) = (1-z) \,, \;\;\;\; I_{qg}^{(1)}(z) = 2z(1-z) \,.
\eea
The parameters going into the anomalous dimensions are listed below. We have
\bea
&&\beta_0 = \frac{11}{3} C_A - \frac{4}{3} T_F n_f  \,,\nn \\
&&\beta_1 = \frac{34}{3} C_A^2 - \frac{20}{3} C_A T_F n_f - 4 C_F T_F n_f \, , \nn \\
&&\beta_2 = \frac{2857}{54}C_A^3 \,
+\left( C_F^2 - \frac{205}{18}C_FC_A - \frac{1415}{54}C_A^2\right)2T_Fn_f \,
+\left( \frac{11}{9}C_F + \frac{79}{54}C_A\right)4T_F^2n_f^2 \,, \nn \\
&&\Gamma_0 = 4 \,,\nn \\
&&\Gamma_1 = 4  \left[ C_A \left( \frac{67}{9} - \frac{\pi^2}{3} \right) -
\frac{20}{9} T_F n_f \right]\,, \nn \\
&&\Gamma_2 = 4\left[
\left( \frac{245}{6}-\frac{134\pi^2}{27}+\frac{11\pi^4}{45}+\frac{22\zeta_3}{3}\right)C_A^2\,
+\left(-\frac{418}{27}+\frac{40\pi^2}{27}-\frac{56\zeta_3}{3} \right)C_AT_Fn_f \right. \nn \\
&&\left.
\quad \quad \quad \quad
+ \left( -\frac{55}{3} + 16\zeta_3\right)C_FT_Fn_f - \frac{16}{27}T_F^2n_f^2
\right]\,, 
\eea
for the $\beta[\alpha_s]$ function and cusp anomalous dimensions. And for the non-cusp 
ones of the hard Wilson coefficient, we have
\bea
\gamma_{H0}^q &= &-6 C_F\,, \nn\\
\gamma_{H1}^q &=& C_F^2 (-3+4 \pi^2-48\zeta_3) \,
+ C_F C_A \left(-\frac{961}{27}-\frac{11\pi^2}{3}+52\zeta_3\right)\nn\\
&& + \, n_f C_F T_F \left(\frac{260}{27}+\frac{4\pi^2}{3} \right)
\,.
\eea
The soft non-cusp anomalous dimensions could be found
in the previous sections, and the anomalous dimension for the beam function
can be obtained through the consistency condition
$\gamma_B^\mu = - \gamma_H^\mu - \frac{1}{2}\gamma_S^\mu$
and $\gamma_B^\nu = -\frac{1}{2}\gamma_S^\nu$ for the normal RG and the SCET
 rapidity evolution, respectively.

Here the $\beta[\alpha_s]$ function is expanded as,
\bea
\beta[\alpha_s] = -2\alpha_s \sum_{n=0} \beta_n \left( \frac{\alpha_s}{4\pi}\right)^{n+1}\,,
\eea
and the rest of the quantities are expanded as
\bea
F[\alpha_s] = \frac{\alpha_s}{4\pi} F_0 + \left(\frac{\alpha_s}{4\pi}\right)^2 F_1 + \cdots \,.
\eea
Note that we can use 
\bea
\mathrm{d} \log \mu = \frac{1}{\beta[\alpha_s]} \mathrm{d}\alpha_s
\eea
to convert the $\log \mu$ integration to $\alpha_s$ integration. 

As for the expansion of $\text{NNLL}^\prime$ resummation,  we need to use
\bea
\frac{1}{\alpha_s(\mu_i)} = 
\frac{X}{\alpha_s(\mu)} + \frac{\beta_1}{4\pi \beta_0} \log(X) \,
+ \frac{\alpha_s(\mu)}{16\pi^2}\left[\,
\frac{\beta_2}{\beta_0}\left(1-\frac{1}{X} \right) \,
+ \frac{\beta_1^2}{\beta_0^2}\left(\,
\frac{\log(X)}{X} + \frac{1}{X} - 1
 \right)
\right]\,,
\eea
with
\bea
X = 1 - \frac{\alpha_s(\mu)}{4\pi}\beta_0 \log\frac{\mu^2}{\mu_i^2} \,.
\eea
For $\mu_i = -\mu - i 0^+$, we have
\bea
X = 1 - ia(\mu) \,,
\eea
with $a(\mu) \equiv \alpha_s(\mu)\beta_0/4$ treated as an ${\cal O}(\alpha_s)$ parameter.


\begin{thebibliography}{99}

\bibitem{AtlasHiggs}
G.~Aad {\it et al.}  [ATLAS Collaboration],
Phys.\ Lett.\ B {\bf 716}, 1 (2012)
[arXiv:1207.7214 [hep-ex]].

\bibitem{CMSHiggs}
S.~Chatrchyan {\it et al.}  [CMS Collaboration],
Phys.\ Lett.\ B {\bf 716}, 30 (2012)
[arXiv:1207.7235 [hep-ex]].

\bibitem{Butterworth:2008iy} 
  J.~M.~Butterworth, A.~R.~Davison, M.~Rubin and G.~P.~Salam,
  Phys.\ Rev.\ Lett.\  {\bf 100}, 242001 (2008)
  [arXiv:0802.2470 [hep-ph]].

\bibitem{TheATLAScollaboration:2013lia} 
  The ATLAS collaboration,
  ATLAS-CONF-2013-079.
\bibitem{Aad:2012gxa} 
  G.~Aad {\it et al.}  [ATLAS Collaboration],
  Phys.\ Lett.\ B {\bf 718}, 369 (2012)
  [arXiv:1207.0210 [hep-ex]].

\bibitem{Chatrchyan:2013zna} 
  S.~Chatrchyan {\it et al.}  [CMS Collaboration],
  arXiv:1310.3687 [hep-ex].
\bibitem{Chatrchyan:2012ww} 
  S.~Chatrchyan {\it et al.}  [CMS Collaboration],
  Phys.\ Lett.\ B {\bf 710}, 284 (2012)
  [arXiv:1202.4195 [hep-ex]].

\bibitem{ATLAS:2013pma} 
  [ATLAS Collaboration],
  ATLAS-CONF-2013-011.
\bibitem{CMS:2013yda} 
  CMS Collaboration [CMS Collaboration],
  CMS-PAS-HIG-13-018.
  
\bibitem{Han:1991ia} 
  T.~Han and S.~Willenbrock,
  Phys.\ Lett.\ B {\bf 273}, 167 (1991).
  
\bibitem{Hamberg:1990np} 
  R.~Hamberg, W.~L.~van Neerven and T.~Matsuura,
  Nucl.\ Phys.\ B {\bf 359}, 343 (1991)
  [Erratum-ibid.\ B {\bf 644}, 403 (2002)].
  
\bibitem{Harlander:2002wh} 
  R.~V.~Harlander and W.~B.~Kilgore,
  Phys.\ Rev.\ Lett.\  {\bf 88}, 201801 (2002)
  [hep-ph/0201206].
  
  
  
  
  

\bibitem{Ferrera:2011bk} 
  G.~Ferrera, M.~Grazzini and F.~Tramontano,
  Phys.\ Rev.\ Lett.\  {\bf 107}, 152003 (2011)
  [arXiv:1107.1164 [hep-ph]].

\bibitem{Dawson:2012gs} 
  S.~Dawson, T.~Han, W.~K.~Lai, A.~K.~Leibovich and I.~Lewis,
  Phys.\ Rev.\ D {\bf 86}, 074007 (2012)
  [arXiv:1207.4207 [hep-ph]].

\bibitem{Shao:2013uba} 
  D.~Y.~Shao, C.~S.~Li and H.~T.~Li,
  arXiv:1309.5015 [hep-ph].


\bibitem{Berger:2010xi} 
  C.~F.~Berger, C.~Marcantonini, I.~W.~Stewart, F.~J.~Tackmann and W.~J.~Waalewijn,
  JHEP {\bf 1104}, 092 (2011)
  [arXiv:1012.4480 [hep-ph]].

\bibitem{Banfi:2012yh} 
  A.~Banfi, G.~P.~Salam and G.~Zanderighi,
  JHEP {\bf 1206}, 159 (2012)
  [arXiv:1203.5773 [hep-ph]].


\bibitem{Banfi:2012jm} 
  A.~Banfi, P.~F.~Monni, G.~P.~Salam and G.~Zanderighi,
  Phys.\ Rev.\ Lett.\  {\bf 109}, 202001 (2012)
  [arXiv:1206.4998 [hep-ph]].

\bibitem{Bauer:2000ew} 
  C.~W.~Bauer, S.~Fleming and M.~E.~Luke,
  Phys.\ Rev.\ D {\bf 63}, 014006 (2000)
  [hep-ph/0005275].

\bibitem{Bauer:2000yr}
C.~W. Bauer, S.~Fleming, D.~Pirjol, and I.~W. Stewart,
\newblock Phys. Rev. {\bf D63}, 114020 (2001), hep-ph/0011336.

\bibitem{Bauer:2001ct} 
  C.~W.~Bauer and I.~W.~Stewart,
  Phys.\ Lett.\ B {\bf 516}, 134 (2001)
  [hep-ph/0107001].

\bibitem{Bauer:2001yt}
C.~W. Bauer, D.~Pirjol, and I.~W. Stewart,
\newblock Phys. Rev. {\bf D65}, 054022 (2002), hep-ph/0109045.

\bibitem{Bauer:2002nz}
C.~W. Bauer, S.~Fleming, D.~Pirjol, I.~Z. Rothstein, and I.~W. Stewart,
\newblock Phys. Rev. {\bf D66}, 014017 (2002), hep-ph/0202088.

\bibitem{Becher:2012qa} 
  T.~Becher and M.~Neubert,
  JHEP {\bf 1207}, 108 (2012)
  [arXiv:1205.3806 [hep-ph]].
  
\bibitem{Becher:2013xia} 
  T.~Becher, M.~Neubert and L.~Rothen,
  arXiv:1307.0025 [hep-ph].


\bibitem{Stewart:2013faa} 
  I.~W.~Stewart, F.~J.~Tackmann, J.~R.~Walsh and S.~Zuberi,
  arXiv:1307.1808 [hep-ph].

\bibitem{Liu:2012sz} 
  X.~Liu and F.~Petriello,
  Phys.\ Rev.\ D {\bf 87}, 014018 (2013)
  [arXiv:1210.1906 [hep-ph]].

\bibitem{Liu:2013hba} 
  X.~Liu and F.~Petriello,
  Phys.\ Rev.\ D {\bf 87}, 094027 (2013)
  [arXiv:1303.4405 [hep-ph]].
  
\bibitem{Boughezal:2013oha} 
  R.~Boughezal, X.~Liu, F.~Petriello, F.~J.~Tackmann and J.~R.~Walsh,
  arXiv:1312.4535 [hep-ph].
  
\bibitem{Alioli:2013hba} 
  S.~Alioli and J.~R.~Walsh,
  arXiv:1311.5234 [hep-ph].

\bibitem{Ligeti:2008ac} 
  Z.~Ligeti, I.~W.~Stewart and F.~J.~Tackmann,
  Phys.\ Rev.\ D {\bf 78}, 114014 (2008)
  [arXiv:0807.1926 [hep-ph]].

\bibitem{Abbate:2010xh} 
  R.~Abbate, M.~Fickinger, A.~H.~Hoang, V.~Mateu and I.~W.~Stewart,
  Phys.\ Rev.\ D {\bf 83}, 074021 (2011)
  [arXiv:1006.3080 [hep-ph]].
  
\bibitem{Gavin:2010az} 
  R.~Gavin, Y.~Li, F.~Petriello and S.~Quackenbush,
  Comput.\ Phys.\ Commun.\  {\bf 182}, 2388 (2011)
  [arXiv:1011.3540 [hep-ph]].

\bibitem{Gavin:2012sy} 
  R.~Gavin, Y.~Li, F.~Petriello and S.~Quackenbush,
  Comput.\ Phys.\ Commun.\  {\bf 184}, 208 (2013)
  [arXiv:1201.5896 [hep-ph]].

\bibitem{Li:2012wna} 
  Y.~Li and F.~Petriello,
  Phys.\ Rev.\ D {\bf 86}, 094034 (2012)
  [arXiv:1208.5967 [hep-ph]].

\bibitem{Brein:2012ne} 
  O.~Brein, R.~V.~Harlander and T.~J.~E.~Zirke,
  Comput.\ Phys.\ Commun.\  {\bf 184}, 998 (2013)
  [arXiv:1210.5347 [hep-ph]].

\bibitem{Campbell:2010ff} 
  J.~M.~Campbell and R.~K.~Ellis,
  Nucl.\ Phys.\ Proc.\ Suppl.\  {\bf 205-206}, 10 (2010)
  [arXiv:1007.3492 [hep-ph]].

\bibitem{Gleisberg:2003xi} 
  T.~Gleisberg, S.~Hoeche, F.~Krauss, A.~Schalicke, S.~Schumann and J.~-C.~Winter,
  JHEP {\bf 0402}, 056 (2004)
  [hep-ph/0311263].
  

\bibitem{Campbell:2013qaa} 
  J.~M.~Campbell, K.~Hatakeyama, J.~Huston, F.~Petriello, J.~Andersen, L.~Barze, H.~Beauchemin and T.~Becher {\it et al.},
  arXiv:1310.5189 [hep-ph].
  
\bibitem{Heinemeyer:2013tqa} 
  SHeinemeyer {\it et al.}  [LHC Higgs Cross Section Working Group Collaboration],
  arXiv:1307.1347 [hep-ph].


\bibitem{Chiu:2011qc} 
  J.~-y.~Chiu, A.~Jain, D.~Neill and I.~Z.~Rothstein,
  Phys.\ Rev.\ Lett.\  {\bf 108}, 151601 (2012)
  [arXiv:1104.0881 [hep-ph]].


\bibitem{Chiu:2012ir} 
  J.~-Y.~Chiu, A.~Jain, D.~Neill and I.~Z.~Rothstein,
  JHEP {\bf 1205}, 084 (2012)
  [arXiv:1202.0814 [hep-ph]].



\bibitem{Ahrens:2008qu} 
  V.~Ahrens, T.~Becher, M.~Neubert and L.~L.~Yang,
  Phys.\ Rev.\ D {\bf 79}, 033013 (2009)
  [arXiv:0808.3008 [hep-ph]].


\bibitem{Martin:2009iq} 
  A.~D.~Martin, W.~J.~Stirling, R.~S.~Thorne and G.~Watt,
  Eur.\ Phys.\ J.\ C {\bf 63}, 189 (2009)
  [arXiv:0901.0002 [hep-ph]].

  

\bibitem{Stewart:2011cf} 
  I.~W.~Stewart and F.~J.~Tackmann,
  Phys.\ Rev.\ D {\bf 85}, 034011 (2012)
  [arXiv:1107.2117 [hep-ph]].

\bibitem{Boughezal:2013uia} 
  R.~Boughezal, F.~Caola, K.~Melnikov, F.~Petriello and M.~Schulze,
  JHEP {\bf 1306}, 072 (2013)
  [arXiv:1302.6216 [hep-ph]].



\bibitem{Becher:2007ty} 
  T.~Becher, M.~Neubert and G.~Xu,
  JHEP {\bf 0807}, 030 (2008)
  [arXiv:0710.0680 [hep-ph]].
  
\bibitem{Ellis:1996nn} 
  R.~K.~Ellis and W.~Vogelsang,
  hep-ph/9602356.
 
  
  
  
\end{thebibliography}
\end{document}